\documentclass[twocolumn,showpacs,preprintnumbers,amsmath,amssymb,pra,superscriptaddress]{revtex4}
\usepackage{amsmath}
\usepackage{graphicx}
\usepackage{dcolumn}
\usepackage{bm}
\usepackage{booktabs}
\usepackage{calc}
\usepackage{multirow}
\usepackage{mathtools}

\usepackage{mathtools}
\usepackage{epsfig}
\usepackage{amssymb}
\usepackage{float}
\usepackage{dcolumn}
\usepackage{bm}
\usepackage{color}
\usepackage{refstyle}

\DeclareMathAlphabet{\mathpzc}{OT1}{pzc}{m}{it}

\newcommand{\vek}[1]{\mathbf{#1}}
\newcommand {\beq} {\begin{eqnarray}}
\newcommand {\eeqn} [1] {\label{#1} \end{eqnarray}}

\newcommand{\rel}[1]{\rho\,}
\newcommand{\ra}[1]{\hat{\mathbf{r}}\,}
\newcommand{\ri}[1]{\mathbf{R}\,}
\newcommand{\ma}[1]{m\,}
\newcommand{\mi}[1]{M\,}
\newcommand{\xa}[1]{x}
\newcommand{\ya}[1]{y\,}
\newcommand{\za}[1]{z\,}
\newcommand{\X}[1]{X\,}
\newcommand{\Y}[1]{Y\,}
\newcommand{\Z}[1]{Z\,}

\newcommand{\toni}[1]{\textcolor{red}{#1}}

\begin{document}

%----------------------------------------------------------------------------------------
%	TITLE
%----------------------------------------------------------------------------------------

\title{Impact of ion motion on atom-ion confinement-induced resonances in hybrid traps}
\date{\today}

%----------------------------------------------------------------------------------------
%	AUTHORS AND AFFILIATIONS
%----------------------------------------------------------------------------------------

\author{Vladimir S.Melezhik}
\email[]{melezhik@theor.jinr.ru} \affiliation{Bogoliubov Laboratory of Theoretical Physics, Joint Institute for Nuclear Research, Dubna, Moscow Region 141980, Russian Federation}
\affiliation{Dubna State University, 19 Universitetskaya Street, Moscow Region 141982, Russian Federation}
\author{Zbigniew Idziaszek}
\email[]{Zbigniew.Idziaszek@fuw.edu.pl}
\affiliation{Faculty of Physics, University of Warsaw, Pasteura 5, 02-093 Warsaw, Poland}
\author{Antonio Negretti}
\email[]{anegrett@physnet.uni-hamburg.de}
\affiliation{Zentrum f\"ur Optische Quantentechnologien, Fachbereich Physik, and Hamburg Center for Ultrafast Imaging, Universit\"at Hamburg, Luruper Chaussee 149, 22761 Hamburg, Germany}

\date{\today}

%----------------------------------------------------------------------------------------
%	ABSTRACT
%----------------------------------------------------------------------------------------

\begin{abstract}
\label{txt:abstract}
We investigate confinement-induced resonances in atom-ion quantum mixtures confined in hybrid traps for small atom-ion mass ratios. Specifically, we consider an ion confined in a time-dependent radio-frequency Paul trap with linear geometry, while the atom is constrained to move into a quasi-one-dimensional optical waveguide within the ion trap. We evaluate the impact of the ion intrinsic micromotion on the resonance position. Thus, we solve the atom-ion dynamics semiclassically, namely the atom dynamics is governed by the three-dimensional time-dependent Schr\"odinger equation, whereas the ion motion is described by the classical Hamilton equations. We find that the energy of the ion provided by the oscillating radiofrequency fields can affect the resonance position substantially. Notwithstanding, the peculiar phenomenology of those resonances regarding perfect transmission and reflection is still observable. These findings indicate that the intrinsic micromotion of the ion is not detrimental for the occurrence of the resonance and that its position can be controlled by the radiofrequency fields. This provides an additional mean for tuning atom-ion interactions in low spatial dimensions. The study represents an important advancement in the scattering physics of compound atomic quantum systems in time-dependent traps.
\end{abstract}

\maketitle

%----------------------------------------------------------------------------------------
%	SECTION 1
%----------------------------------------------------------------------------------------

\section{INTRODUCTION}

Compound atom-ion systems afford a new platform to study quantum physics in which multi-energy and multi-length scales are involved. In particular, atom-ion systems allow investigating condensed-matter systems more closely. For instance, an important component of a solid-state system is the electron-phonon coupling, which is mimicked naturally in an atom-ion system. Indeed, the atom-ion interaction has the effect that the passage of an atom in the proximity of an ion crystal influences the state of the ions via the exchange of phonons as in a real solid-state system. This feature is absent in ultracold atoms trapped in an optical lattice, where there is no back-action of the atoms on the lattice. Furthermore, with such compound system it is also possible to study the formation of mesoscopic molecular ions~\cite{SchurerPRL17,CotePRL02} and charge transport~\cite{CotePRL00,MukherjeePRA19}, to mention a few examples (for a detailed overview, see Ref.~\cite{Tomza}). The realisation of compound atom-ion systems in the laboratory, however, is quite challenging, since it combines different trap technologies such as radiofrequency traps for ions and optical dipole traps for atoms and this limits fundamentally the attainable temperatures.
Albeit the s-wave collisional quantum regime (i.e. the s-wave regime) has been not yet reached in current atom-ion laboratories, a significant experimental effort has been put forward in very recent years. Specifically, it has been shown that sub-microkelvin temperatures can be attained when ionising a Rydberg atom in a Bose-Einstein condensate~\cite{KleinbachPRL18} and that the required collisional energies to enter the s-wave limit in radiofrequency traps are attained when choosing a small atom-ion mass ratio~\cite{CetinaPRL12,NguyenPRA12,JogerPRA14,TomzaPRA15,FuerstJPB18}, which is within experimental reach~\cite{Feldker2019}. Furthermore, proof-of-principle experiments have demonstrated laser-controlled atom-ion interactions~\cite{EngelPRL18,EwaldPRL19,Haze2019}, therefore opening new possibilities for controlling interactions and developing light-matter interfaces for quantum information processing~\cite{SeckerPRA16,SeckerPRL17}.
We underscore that the attainment of the s-wave regime is crucial for the observation of atom-ion Feshbach resonances~\cite{IdziaszekPRA09} as well as for quantum technological applications with atom-ion systems such as quantum gates~\cite{DoerkPRA10}, quantum simulation of the electron-phonon coupling~\cite{BissbortPRL13}, and for reaching the strong-coupling polaron regime~\cite{CasteelsJLTP11}.

Confinement-induced resonances (CIRs)~\cite{Olshanii1998,KimMelSch,MelKimSch,MelezhikPRA16,MartaPRA18,MelShad} have been pivotal in entering the regime of strongly correlated atomic matter~\cite{HallerScience09}. A CIR occurs in an atomic trap  when the atom-atom scattering length in free space, $ a_{s}$, becomes comparable to the transversal width, $a_{\perp}$, of the trap: $a_{\perp} / a_{s} \rightarrow $ 1.4603~\cite{Olshanii1998}. At that ratio the bound state in the closed channel, that is, the first excited state of the transverse confinement, coincides with the collision threshold of the entrance channel. In this case, the scattering amplitude $f(a_{\perp}, a_{s})$ of the atom on atom tends to -1, and therefore the transmission $T = \mid 1 + f \mid^2$ approaches zero \cite{MelezhikPRA16,Olshanii1998,KimMelSch,MelKimSch}. Thus, by varying $a_{\perp}$ and $a_{s}$ near the CIR, it is possible to control the effective atom-atom interaction of the confined atomic system. As for the neutral atomic counterpart, atom-ion or atom-dipole CIRs can be an additional `knob' for manipulating the mixture's interaction in low spatial dimensions. For instance, CIRs can be utilised to steer the particle flow in Josephson junctions~\cite{JogerPRA14,SchurerPRA16,Ebgha2019}, to tune interactions in a Tomonaga-Luttinger liquid in which a linear ion crystal is immersed~\cite{Michelsen2019}, for precise magnetometry~\cite{JachymskiPRL18,WasakPRA18}, and quantum simulation~\cite{PupilloPRL08,BissbortPRL13,NegrettiPRB14,DehkharghaniPRA17,CuadraPRL18,CuadraPRB19}.
This is especially important in the atom-ion setting when state-dependent atom-ion interactions are needed to perform particular quantum information processing tasks. Indeed, experiments~\cite{RatschbacherPRL13,FuerstPRA18,SikorskyPRL18} have shown that spin-exchange collisions can occur after a few Langevin collisions because of spin-orbit couplings~\cite{TscherbulPRL16}. This effect has a minor impact in a quasi-one dimensional geometry.

A first theoretical study investigating the possibility of realising CIRs in an atom-ion system was done in Ref.~\cite{IdziaszekPRA07} for the case of time-independent atom and ion traps with identical frequencies. In Ref.~\cite{MelezhikPRA16} the CIRs in atom-ion hybrid systems were predicted and the conditions for the atom-ion CIR appearance were obtained in the static ion approximation, where the ion is pinned in a precise position in space and cannot move. Interestingly, it has been found an isotope-like effect of the resonance position. While in the neutral setting the resonance condition is attained when $a_{\perp} / a_{s} \rightarrow $ 1.4603, as previously discussed, in atom-ion systems, when the effective spatial range, $R^*$, of the polarisation potential, $-C_4/r^4$, is comparable to $a_{\perp}$ (or even larger), the position of the CIR strongly relies on the atom-ion mass ratio~\cite{MelezhikPRA16}.
%Here $a_{\perp}$ denotes the waveguide width and $a_{s}$ the atom-atom or atom-ion three-dimensional s-wave scattering length in free space. This effect is much harder to observe in cold atomic systems, as background scattering lengths are typically on the order of a few nanometers, that is, the interaction is very much short-ranged, while in atom-ion mixtures scattering lengths can assume values up to a few hundreds of nanometers. Hence, it renders the observation of such mass reliance on the atom-ion CIR much easier to accomplish in the laboratory.
This effect is much harder to observe in neutral atomic systems, as it requires large trap frequencies (hundreds of kHz) in order to produce small trap widths $a_{\perp}$, while atom-ion systems require frequencies $\le 100$ kHz~\cite{MelezhikPRA16}.

In view of the foregoing, it becomes relevant to go beyond the static approximation in the problem of atom-ion CIR. So far, however, atom-ion collisions in radiofrequency traps have been treated either purely classically~\cite{CetinaPRL12} or within the Markovian quantum master equation formalism~\cite{KrychPRA15}. Both studies indicated that the most favourable atom-ion species for reaching the quantum regime is Li/Yb$^+$. In the study of Ref.~\cite{KrychPRA15}, however, the atom-ion interaction has been treated perturbatively, i.e. in the Born approximation, which omits, for instance, the resonance effects in the atom-ion scattering.
In this work we go one step further in the quantum mechanical treatment of atom-ion collisions in Paul traps and extend the previous work of two of us~\cite{MelezhikPRA16}. Indeed, we treat the atom dynamics fully quantum mechanically with the time-dependent Schr\"odinger equation, whereas the ion motion is described classically via the Hamilton equations. Such a semiclassical treatment is well justified when the ion is much heavier than the colliding atom. As we explain in detail below, the equations of motion of the atom and the ion are coupled via the atom-ion interaction. In the atom Schr\"odinger equation the ion position is treated as a time-dependent parameter, while the interaction in the ion Hamilton equations is averaged quantum mechanically over the instantaneous quantum state of the atom.

Towards this aim, the quantum-semiclassical computational method~\cite{MelSchm,Melezhik2001,MelezhikCohen,MelSev} specifically designed for particle collisions such as the problem of ionisation of the helium ion colliding with protons~\cite{MelezhikCohen} and antiprotons~\cite{MelSev} has been employed and  extended to the time-dependent domain, as our radiofrequency ionic confinement requires. Moreover,
our analyses focus on the specific Li/Yb$^+$ atom-ion pair, since it is the most promising atomic pair to reach the s-wave regime in Paul traps and it is currently under intense experimental investigations~\cite{JogerPRA17,FuerstPRA18,Feldker2019}. We note, however, that with regard to the mass ratio a good atom-ion pair is also Li/Ca$^+$~\cite{HazePRA15,SaitoPRA17,HazePRL18}. Our analysis shows that the intrinsic micromotion of the ion, which is unavoidable in Paul traps (differently from optical trapping of ions~\cite{SchmidtPRX18}), is not detrimental for the occurrence of an atom-ion CIR. We find that the CIR position strongly relies on the ion kinetic energy. This implies that the atom-ion interaction can be controlled not only by the short-range atom-ion physics or by means of magnetic Fano-Feshbach resonances as well as the width of the atomic waveguide, but also by the ion kinetic energy, which can be manipulated via the external driving, e.g., by changing the radiofrequency of the Paul trap.

The paper is organised as follow. In Sec.~\ref{sec:Hamiltonian} we introduce our microscopic atom-ion Hamiltonian and the corresponding equations of motion, namely the coupled Schr\"odinger-Hamilton equations. Beside this, we discuss how to determine the scattering amplitude in the time-dependent scenario. In Sec.~\ref{sec:results} we present our findings and discuss the physical implications. In Sec.~\ref{sec:conclusions} we draw our conclusions and provide an outlook for future work.

%----------------------------------------------------------------------------------------
%	SECTION 2
%----------------------------------------------------------------------------------------

\section{Problem formulation and methodology}
\label{sec:Hamiltonian}

\begin{figure}
\centering\includegraphics[scale=0.4]{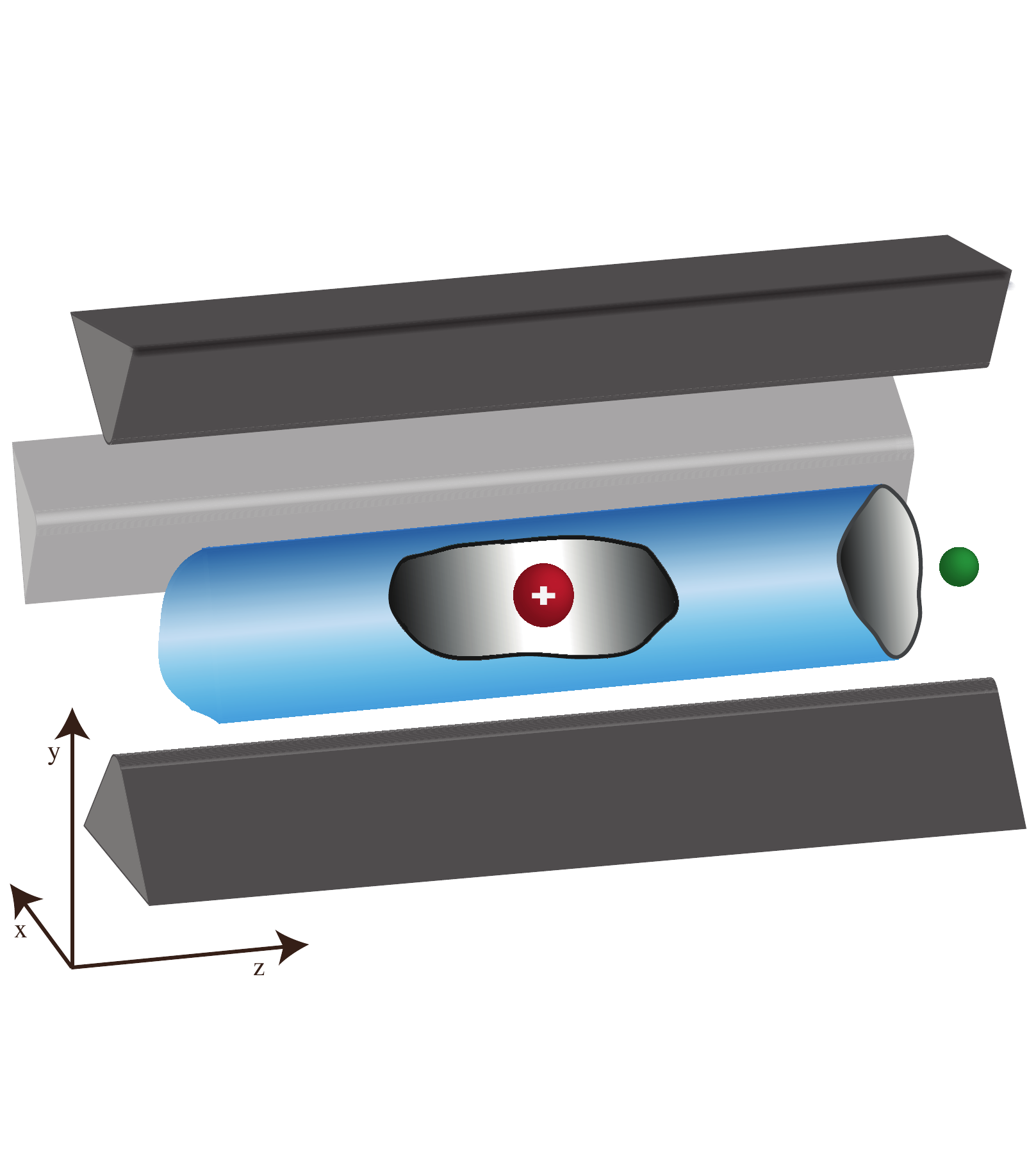}
\caption{ (color online) Pictorial illustration of the system under investigation in this study. The light- and dark-grey electrodes (the big bars in the figure) of the Paul trap generate the time-dependent electric fields needed to confine the ion (red sphere) transversally, whereas longitudinally a static voltage is applied to ensure confinement (not shown). The atom (green sphere) is injected from the right to the left into a waveguide (light-blue), whose centre hosts the ion. The waveguide is orientated along the longitudinal axis, $z$, of the linear Paul trap. In the transverse directions, $x,y$, the confining potential both for the atom and the ion is strong.}
\label{fig:sketch}
\end{figure}

In this section we describe theoretically the atom-ion system including the equations of motion for the atom and the ion, how do we face the scattering problem in time-dependent radiofrequency traps, and what are the scattering quantities of interest in order to assess the CIR position in reliance of the ion kinetic energy and atomic waveguide width. A schematic view of the system under investigation is given in Fig.~\ref{fig:sketch}.

%----------------------------------------------------------------------------------------
%	SUBSECTION 2.1
%----------------------------------------------------------------------------------------

\subsection{The atom quantum Hamiltonian}

The atom is described by the Hamiltonian

\begin{align}
\label{eq:Ha}
\hat H_a(\hat{\vek{r}}_a, t;\vek{r}_i) =
-\frac{\hbar^2}{2 m_a}\nabla_a^2 + \frac{m_a\omega_{\perp}^2}{2} (\hat x_a^2+\hat y_a^2)\nonumber\\
+ V_{ai}(\vert\hat{\vek{r}}_a-\vek{r}_i(t)\vert),
\end{align}
where $\hat{\vek{r}}_a\equiv (\hat x_a,\hat y_a,\hat z_a)$ and $\vek{r}_i\equiv (x_i,y_i,z_i)$ are the atom position operator and ion position vector, respectively, $m_a$ is the atom mass, and $\nabla^2_a \equiv \partial_{x_a^2}+\partial_{y_a^2}+\partial_{z_a^2}$ is the laplacian operator ($\partial_{x_a}$ denotes the partial derivative, for instance, in the $x_a$ direction). The second term in Eq.~(\ref{eq:Ha}) describes the harmonic potential of frequency $\omega_{\perp}$ due to the waveguide, whereas along the longitudinal direction, $z_a$, the atom does not experience any confinement.
The transverse potential is assumed to be so tight to render the atom motion quasi-one-dimensional (quasi-1D).
We note that in the Hamiltonian~(\ref{eq:Ha}) we explicitly introduced the parametric dependence on the ion position.

The last term in Eq.~(\ref{eq:Ha}), $V_{ai}(\vert\hat{\vek{r}}_a-\vek{r}_i(t)\vert)$, describes the interaction between the atom and ion,
whose asymptotic behaviour at large atom-ion separations $r(t)\equiv\vert \hat{\vek{r}}_a - \vek{r}_i(t)\vert \rightarrow \infty$ has the form
\begin{align}
\label{eq:Vai}
V_{ai}(r(t)) \simeq -\frac{C_4}{r(t)^4}\,,
\end{align}
i.e. it behaves as the polarisation potential. The dispersion coefficient is given by $C_{4} = \frac{\alpha e^2}{2}\frac{1}{4\pi\epsilon_0}$ (in SI units) with $\alpha$ being the static atom polarisability, $e$ the electron charge, and $\epsilon_0$
the vacuum permittivity.
At short-range distances, when the two electronic clouds do overlap, the potential is repulsive, albeit its form is generally unknown.
For the sake of numerical convenience, we use the regularised potential~\cite{KrychPRA15}
\begin{align}
\label{eq:Vaireg}
V_{ai}(r(t)) =-\frac{[r^2(t)-c^2]}{[r^2(t)+c^2]}\frac{C_4}{[r^2(t)+b^2]^2}\,.
\end{align}
This potential reproduces at large distances the polarisation potential~(\ref{eq:Vai}), whereas at $r=0$ it assumes a large, but finite, numerical value, contrarily to the typically employed $C_{12}/r^{12}$ potential.
Variation of the two parameters $b$ and $c$ permits to tune the atom-ion interaction for any value $-\infty <a_{s}<+\infty$ of the atom-ion scattering length in free space in the zero-energy limit.
We note that the time-dependence of the Hamiltonian~(\ref{eq:Ha}) enters via the ion trajectory $\vek{r}_i(t)$ in the atom-ion interaction~(\ref{eq:Vaireg}). The ion trajectory is obtained by simultaneous solving the classical equations of motion for a charge particle in a Paul trap, as explained in the next section.

%----------------------------------------------------------------------------------------
%	SUBSECTION 2.2
%----------------------------------------------------------------------------------------

\subsection{The ion classical Hamiltonian}
\label{sec:ionH}

The ion is assumed to be confined in a linear Paul trap, whose electric fields read as~\cite{LeibfriedRMP03}
\begin{align}
\vek{E}_{\mathrm{s}} &= \frac{m_i}{2\vert e\vert} \omega_i^2 \left(x_i,y_i, -2z_i\right),\nonumber\\
\vek{E}_{\mathrm{rf}} &= \frac{m_i\Omega_{rf}^2 q}{2\vert e\vert} \cos(\Omega_{rf} t)\left(x_i,-y_i, 0\right)\,.
\end{align}
Here, $m_i$ is the ion mass, $\Omega_{rf}$ is the radiofrequency (rf), $\omega_i = \Omega_{rf}\sqrt{a/2}$,
$q$ and $a$ are dimensionless geometric parameters (i.e. $q_z = 0$, $q_y = - q_x\equiv q$, $-a_z/2 = a_x = a_y \equiv a$, and $a\ll q^2 <1$). Hereafter, we assume that the axis of the waveguide in which is travelling the colliding atom is precisely the $z$-axis of the Paul trap (see also Fig.~\ref{fig:sketch}). The corresponding
non-conservative potential is given by
\begin{align}
\label{eq:Uion}
U(\vek{r}_i,t) &= \frac{m_i\omega_i^2}{2}\left(z_i^2-\frac{x_i^2+y_i^2}{2}\right) \nonumber\\
\phantom{=}&+\frac{m_i\Omega_{rf}^2}{2} q \cos(\Omega_{rf} t)\left(\frac{y_i^2}{2}-\frac{x_i^2}{2}\right)\,.
\end{align}
Hence, the classical Hamiltonian describing an ion in a Paul trap is given by
\begin{align}
H_i^{trap}(\vek{p}_i,\vek{r}_i,t) = \frac{\vek{p}_i^2}{2 m_i} + U(\vek{r}_i,t)\,.
\end{align}

%----------------------------------------------------------------------------------------
%	SUBSECTION 2.3
%----------------------------------------------------------------------------------------

\subsection{The atom-ion equations of motion}
\label{sec:EOM}

When the atom is confined in the waveguide within the Paul trap, the ion will experience its presence via the atom-ion interaction modelled by Eq.~(\ref{eq:Vaireg}). The full classical ion Hamiltonian is therefore given by
\begin{align}
H_i(\vek{p}_i,\vek{r}_i,t;\vek{r}_a) = H_i^{trap}(\vek{p}_i,\vek{r}_i,t)\nonumber\\
+ \langle V_{ai}(\vert\hat{\vek{r}}_a-\vek{r}_i(t)\vert)\rangle \,,
\end{align}
where
\begin{align}
	\langle V_{ai}(\vert\hat{\vek{r}}_a-\vek{r}_i(t)\vert)\rangle = \nonumber\\ \langle\Psi(\vek{r}_a,t;\vek{r}_i)\vert V_{ai}(\vert\hat{\vek{r}}_a-\vek{r}_i(t)\vert)\vert\Psi(\vek{r}_a,t;\vek{r}_i)\rangle
\end{align}
is the quantum mechanical average of the atom-ion interaction. As for the atom case, also for the ion Hamiltonian we explicitly emphasise the parametric dependence on the atom position.

The atom wavefunction $\Psi(\vek{r}_A,t)$ is governed by the time-dependent Schr\"odinger equation
\begin{align}
\label{eq:SE}
i\hbar\frac{\partial}{\partial t}\Psi(\vek{r}_a,t;\vek{r}_i) = \hat H_a(\hat{\vek{r}}_a,t;\vek{r}_i) \Psi(\vek{r}_a,t;\vek{r}_i)\,,
\end{align}
where $\hat H_a(\hat{\vek{r}}_a,t;\vek{r}_i)$ is defined by Eq.~(\ref{eq:Ha}).
Hence, the dynamics of the ion is governed by the Hamilton equations
\begin{align}
\label{eq:Hamilton}
\frac{d}{d t}\vek{p}_i & = -\frac{\partial}{\partial \vek{r}_i}H_i (\vek{p}_i,\vek{r}_i,t;\vek{r}_a)\,,\nonumber\\
\frac{d}{d t}\vek{r}_i & = \frac{\partial}{\partial \vek{p}_i}H_i (\vek{p}_i,\vek{r}_i,t;\vek{r}_a)\,.
\end{align}
This set of classical equations together with the atom Schr\"odinger equation~(\ref{eq:SE}) forms the complete set of dynamical equations for describing the confined atom-ion collision in hybrid traps.
In the present study we consider collisions of a light atom with a much heavier ion in the range of very low atomic colliding energies $E_{\mathrm{coll}}$
(ultracold atoms), where the relation $p_a=\sqrt{2m_a E_{\mathrm{coll}}} \ll p_i$ for their momentums is satisfied.
In addition,  we require that $E_i = p_i^2/(2 m_i) \gg \hbar \omega_i$, which further justifies the application of the classical description for the ion.
Additionally, because of the separation of energy (i.e. time) scales among the ion and atom dynamics, in the Hamilton equations~(\ref{eq:Hamilton}) we neglect the functional derivatives owed to the parametric reliance of the atom wavefunction on the ion position, that is,
\begin{align}
\label{eq:funcDer}
\frac{\partial}{\partial \xi_i}\Psi(\vek{r}_a,t;\vek{r}_i) \equiv 0 \qquad \xi_i=x_i,\,y_i,\,z_i.
\end{align}

In order to integrate simultaneously the equations~(\ref{eq:SE}) and~(\ref{eq:Hamilton}), we need proper initial conditions with physical significance. At the beginning of the collisional process, the atom and the ion are assumed to be far away from each other such that they do not interact ($V_{ai}=0$). In particular, the atom is initially in the ground state of the atomic trap with the low longitudinal colliding energy, that is, $E_{\mathrm{coll}} \ll 2\hbar\omega_{\perp}$, whereas the ion performs fast (with respect of atom motion) oscillations in the Paul trap with mean transversal $\bar{E}_{\perp}$ and longitudinal $\bar{E}_{\parallel}$ energies. Since the atom approaches the region of interaction with the ion very slowly ($E_{\mathrm{coll}} /\hbar \ll \omega_{\perp} \ll \omega_i,\Omega_{rf}$), the initial position of the ion does not influence the scattering process itself, which depends only on $\bar{E}_{\perp}$ and $\bar{E}_{\parallel}$. Specifically, the classical solution of the ion equations of motion (Mathieu equation) in the Paul trap (without the atom) are well approximated by: $A_j \cos(\omega_i t + \phi_j) [1 + q_j\cos(\Omega_{rf} t)/2]$ $\forall\,j=x,y,z$~\cite{BerkelandJAP98}. The associated kinetic energy depends on the amplitude $A_j$, but not on the phase $\phi_j$~\footnote{We have verified this fact by numerical simulations for our parameters as well.}. Therefore, we choose, without loss of generality, the ion position at the initial time $t=0$ in the trap centre with transversal energy,
$E_{\perp}$, and longitudinal energy, $E_{\|}$. This can be summarised with the following set of initial conditions:
\begin{align}
\label{eq:initialion}
\vek{r}_i(t=0) &= (0,0,0),\nonumber\\
p_{i,x}(t=0) &= \sqrt{2 m_i E_{\perp}},\nonumber\\
p_{i,y}(t=0) &= 0,\nonumber\\
p_{i,z}(t=0) &= \sqrt{2 m_i E_{\parallel}}.
\end{align}
These initial conditions set the mean values of the ion transversal and longitudinal energies as $\bar{E}_{\perp}=1.64 E_{\perp}$ (calculated numerically for an ytterbium ion in a trap with $\Omega_{rf}=2\pi\times 2$MHz, $\omega_i = 2\pi\times 63$kHz, $a=0.002$ and $q=0.08$) and $\bar{E}_{\parallel}= E_{\parallel}/2$, which is in qualitative agreement with the estimate
\begin{align}
\bar{E}_{\perp}=\frac{E_{\perp}}{2}\left[1+\left(\frac{q\Omega_{rf}}{2\omega_i}\right)^2\right] \simeq 1.3 E_{\perp}
\end{align}
from the first-order solution of the Mathieu equation~\cite{Landau,BerkelandJAP98}.
We also note that such a choice for the initial transversal momentum orientation, that is, along the $x$-axis, is not relevant for the scattering problem we are interested in. Indeed, as we have verified in our numerical simulations, the final result is invariant relatively to the initial orientation of the ion transversal momentum $\vek{p}_{i,\perp}(t=0)$ in the $x-y$ plane. This fact is also a consequence of the cylindrical symmetry of both the atomic waveguide and the linear Paul trap. Hence, we can safely assume that along the $y$-axis the initial energy is zero.

As far as the initial condition for the atom wavefunction is concerned, we use the following ansatz~\cite{MelKimSch}
\begin{align}
\label{eq:initialatom}
\psi(\vek{r}_a,t=0)& =N \varphi_{0}(\rho_a)
e^{-\frac{(z_a-z_{0})^{2}}{2a_{z}^{2}}} e^{ikz_a} \nonumber\\
\phantom{=} & = N \varphi_{0}(\rho_a)\chi(z_a-z_0)e^{ikz_a}\,.
\end{align}
Here $N$ is a normalisation constant, $\rho_a=r_a \sin\theta_a$ with $r_a\geq 0$ and \toni{$\theta_a\in [0,\pi)$}, $\hbar k=\sqrt{2m_aE_{coll}}$ is the initial momentum, $z_0$ is the position of the initial wave-packet $\chi$, which is far from the ion location, that is, at $z_a = z_0$ the atom and ion do not interact: $V_{ai}(z_a=z_0,r_i)\rightarrow 0$. The longitudinal width of the initial wave-packet (\ref{eq:initialatom}) is
chosen sufficiently broad according to $a_z \simeq 30-40 R^*$ ($R^*=\sqrt{2\mu C_4}/\hbar$ is the characteristic length scale of the atom-ion interaction) to satisfy the demand of sufficient monochromaticity of the
wave-packet along the $z$ direction $D_k(t=0)=\langle\psi(t=0)\vert(k-\bar{k)^2}\vert\psi(t=0)\rangle\rightarrow 0$. This choice of $a_z$ provides insignificant  deformation of the envelop $\chi$ during scattering due to the small dispersion $D_k(t)$ of the wave-packet~\cite{MelKimSch}. The above initial condition is interpreted as follows: Initially, the atom and ion are far from each other such that the atom is initially prepared in the ground state of the transverse confinement, that is, in the ground state $\varphi_{0}$ of the two-dimensional harmonic oscillator as well as in the ground state of an optical dipole trap that is approximated by a harmonic potential of frequency $\hbar/(m_a a_z^2)$. At times $t>0$ the atomic longitudinal confinement is suddenly switched off and an initial momentum kick is imparted to the atom (e.g., via a Raman configuration of lasers) such that the atom wave-packet is moving towards the ion trap centre with velocity $v_{0}=\hbar k/m_a$.

For details on the numerical implementation of the integrators of the atom-ion equations of motion, we refer the interested reader to the appendix~\ref{sec:numerics}.

%----------------------------------------------------------------------------------------
%	SUBSECTION 2.4
%----------------------------------------------------------------------------------------

\subsection{Determination of the scattering amplitude}
\label{sec:scattering}

In the course of the collisional process the atom wave-packet splits
up into two parts each of them moving in opposite directions
$z_a\rightarrow \pm\infty$. Asymptotically we encounter the
following behaviour~\cite{Olshanii1998,Kim,MelKimSch}
\begin{widetext}
\begin{align}
\label{eq:ansatz-withion}
\psi(\rho_a,z_a,t\rightarrow +\infty)
&\mathop{\longrightarrow}\limits_{z_a \rightarrow
+\infty}\psi^{+}(\rho_a,z_a,t) =(1+f^{+}(k)) N
\varphi_{0}(\rho_a)\tilde{\chi}(z_a-(z_{0}+vt))
e^{ik_fz_a}, \nonumber\\
\psi(\rho_a,z_a,t\rightarrow +\infty)
&\mathop{\longrightarrow}\limits_{z_a \rightarrow
-\infty}\psi^{-}(\rho_a,z_a,t) = f^{-}(k) N
\varphi_{0}(\rho_a) \tilde{\chi}(-z_a-(z_{0}+vt))
e^{-ik_fz_a}\,.
\end{align}
\end{widetext}
Here $f^{\pm}(k)$ are the atom-ion forward and backward scattering amplitudes in
the presence of the external confining potential due to the atomic waveguide and the ion radiofrequency trap.
The function $\tilde{\chi}(z_A,t)$ describes the atom motion in the longitudinal $z$-direction, namely the spreading
of the initial Gaussian wave-packet.

Our goal is to determine the forward scattering amplitude $f^{+}(k)$.
Towards that end, we solve first the atom Schr\"odinger equation without the atom-ion interaction, but in the confining waveguide.
The corresponding solution at $z_a \rightarrow +\infty$ is
\begin{align}
\label{eq:ansatz-withoution}
\psi^{(0)+}(\rho_a,z_a) = N \varphi_{0}(\rho_a)\,\tilde{\chi}(z_a-(z_{0}+vt))\, e^{ikz_a}\,,
\end{align}
whereas at distances $z_a \rightarrow -\infty$ it is identically zero, that is, $\psi^{(0)-}(\rho_a,z_a) \equiv 0$, as a consequence of the fact that there is no scattering centre, i.e. no ion. Because of the sufficient monochromaticity of the wave-packet (see the set of the envelop $\chi$ in the initial condition (\ref{eq:initialatom})) and the unitarity of the Schr\"odinger equation, it holds
\begin{align}
\label{eq:overlap}
\langle\psi^{(0)+}(t)\vert\psi(t)\rangle\mathop{\longrightarrow}\limits_{t
\rightarrow +\infty}1+f^{+}(k)\,.
\end{align}
This relation is used to calculate the amplitude $f^{+}(k)$. In order to arrive to the result on the right-hand-side of Eq.~(\ref{eq:overlap}), we performed the approximation $e^{i(k-k_f)z_a}\simeq 1$ within the region $\vert z_a-(z_0+vt)\vert \leq a_z$, where the wave packets~(\ref{eq:ansatz-withion}) and~(\ref{eq:ansatz-withoution}) do overlap, as a consequence of the smallness of the transmitted momentum during the collision [i.e. $2\pi/(k-k_f)$ is much larger than the width of $\tilde{\chi}$].
This approximation has been verified numerically by evaluating the mean atomic energy after the collision
$\bar{E}_{a}(t\rightarrow +\infty) = \mathop{\lim}\limits_{t\rightarrow +\infty}\langle\psi(t)\vert\hat H_a\vert\psi(t)\rangle$, which did not exceed the threshold of transverse atomic excitations $3\hbar\omega_{\perp}$.

By exploiting the current conservation law, the transmission, $T$, and reflection, $R$, coefficients are defined as
\begin{align}
T(k) = | 1+ f^+(k)|^2,\,\,\,\,\,\,\,\,\, R(k) = 1 - | 1+ f^+(k)|^2\,.
\end{align}
We note that the above expression for the reflection coefficient, $R(k)$, holds for elastic scattering processes only.
For inelastic scattering the current conservation law is in general violated and one has to replace
the reflection coefficient by the expression $R=|f^-(k)|^2$. This occurs, for instance, when the atom energy is above the energy of the transverse excited state, that is, larger than $3\hbar\omega_{\perp}$. In our study, however, these processes are not treated.

Finally, in the $s$-wave zero-energy limit $f^{+}=f^{-}$, the quasi-1D coupling constant is given by~\cite{Olshanii1998,Berg}
\begin{align}
\label{eq:g1D}
g_{1D}=\mathop{\lim}\limits_{k
\rightarrow 0}
\frac{\hbar^2 k}{m_a}\frac{\Re\{f^{+}(k)\}}{\Im\{f^{+}(k)\}}.
\end{align}
The coupling constant $g_{1D}$ is the most relevant parameter for analysing confined scattering close to a CIR, where $g_{1D}\rightarrow \pm\infty$~\cite{Olshanii1998,KimMelSch,MelKimSch,MelShad}. However, simultaneously with $g_{1D}$, the position of the CIR is also controlled by checking the condition $T\rightarrow 0$.

The above procedure for determining the scattering parameters has been already utilised for ultracold atomic collisions in a waveguide-like atomic trap~\cite{MelKimSch}. The high numerical accuracy of the calculation obtained with this procedure for the transmission and reflection coefficients in wide range of the interactions including the resonant region of the CIR appearance was confirmed by a comparison with simple estimates obtained from direct calculations of the atomic probability to be out of the scattering center after the collision.

%----------------------------------------------------------------------------------------
%	SECTION 3
%----------------------------------------------------------------------------------------

\section{Results}
\label{sec:results}

We here investigate in detail the scattering dynamics of a ytterbium ion confined in a linear Paul trap and a lithium atom in a waveguide. Atom-ion experiments with such an atomic pair are currently investigated intensively~\cite{JogerPRA17,FuerstPRA18,Feldker2019}.
The atomic trap frequency, $\omega_{\perp}$, has been chosen within the range $2\pi \times(1 - 11)$kHz, whereas for the Paul trap parameters we have chosen:
$\Omega_{rf}=2\pi\times 2$ MHz, $\omega_i = 2\pi\times 63$ kHz, $a=0.002$ and $q=0.08$ [see also Eq.~(\ref{eq:Uion})]. All numerical simulations have been performed with initial longitudinal atom energy $E_{\|}/k_B=m_a v_0^2/(2k_B)$ from the region $\sim$ 10 nK of low energies.

\begin{figure*}%[H]
\parbox{0.95\textwidth}{
\includegraphics[width=0.45\textwidth,height=0.45\textwidth]{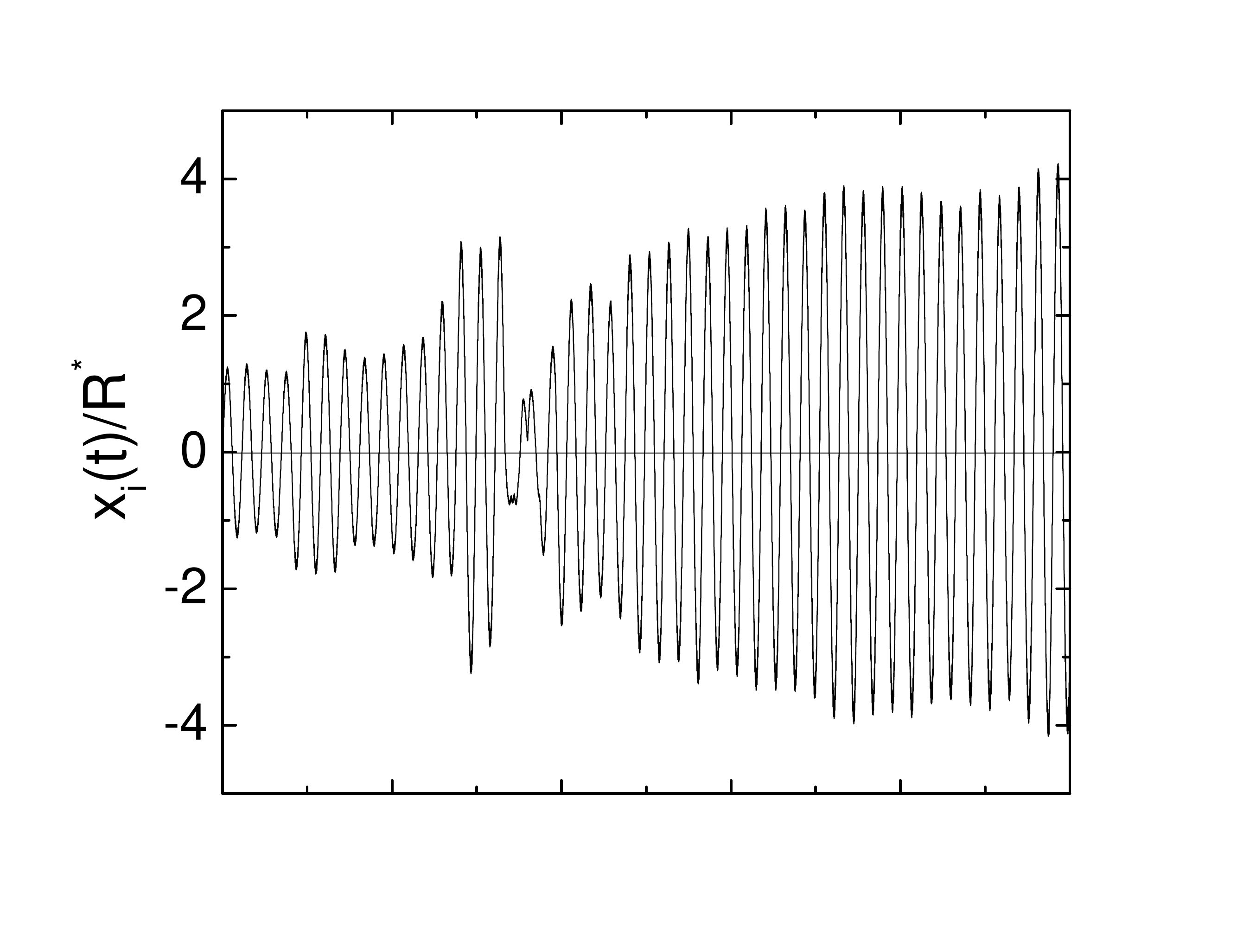}
\includegraphics[width=0.40\textwidth,height=0.40\textwidth]{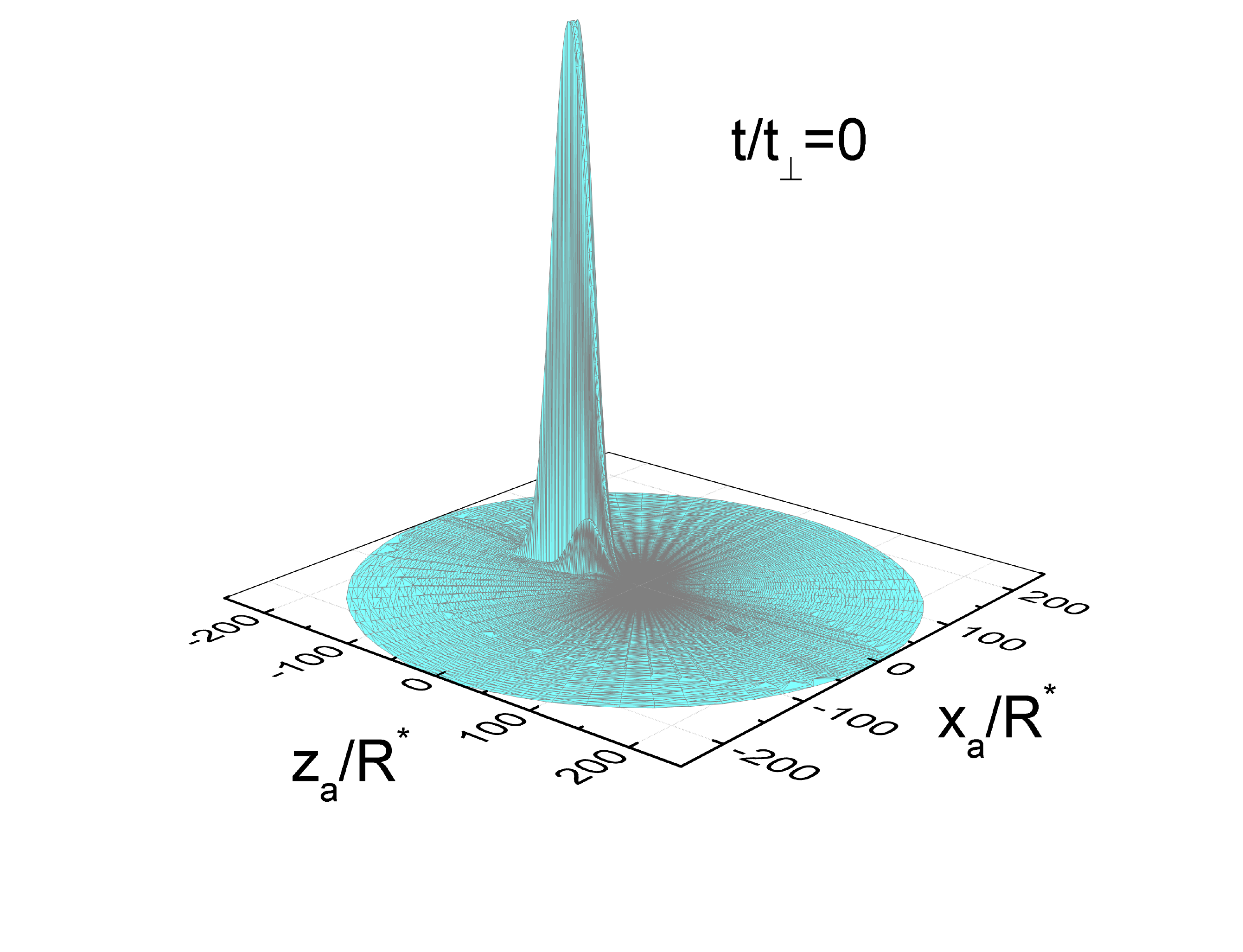}
\vspace{-1.2cm}}
\parbox{0.95\textwidth}{
\includegraphics[width=0.45\textwidth,height=0.45\textwidth]{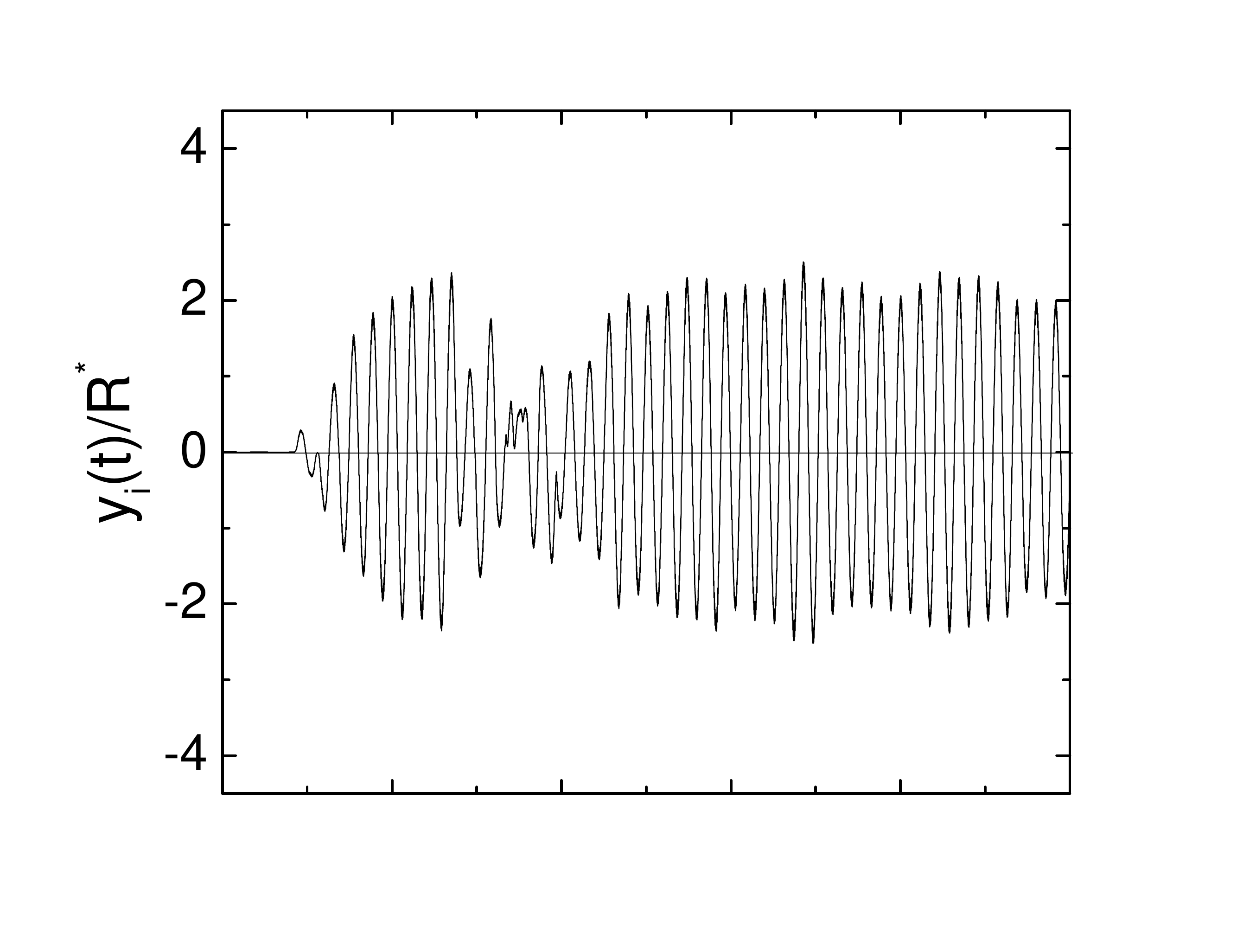}
\includegraphics[width=0.40\textwidth,height=0.4\textwidth]{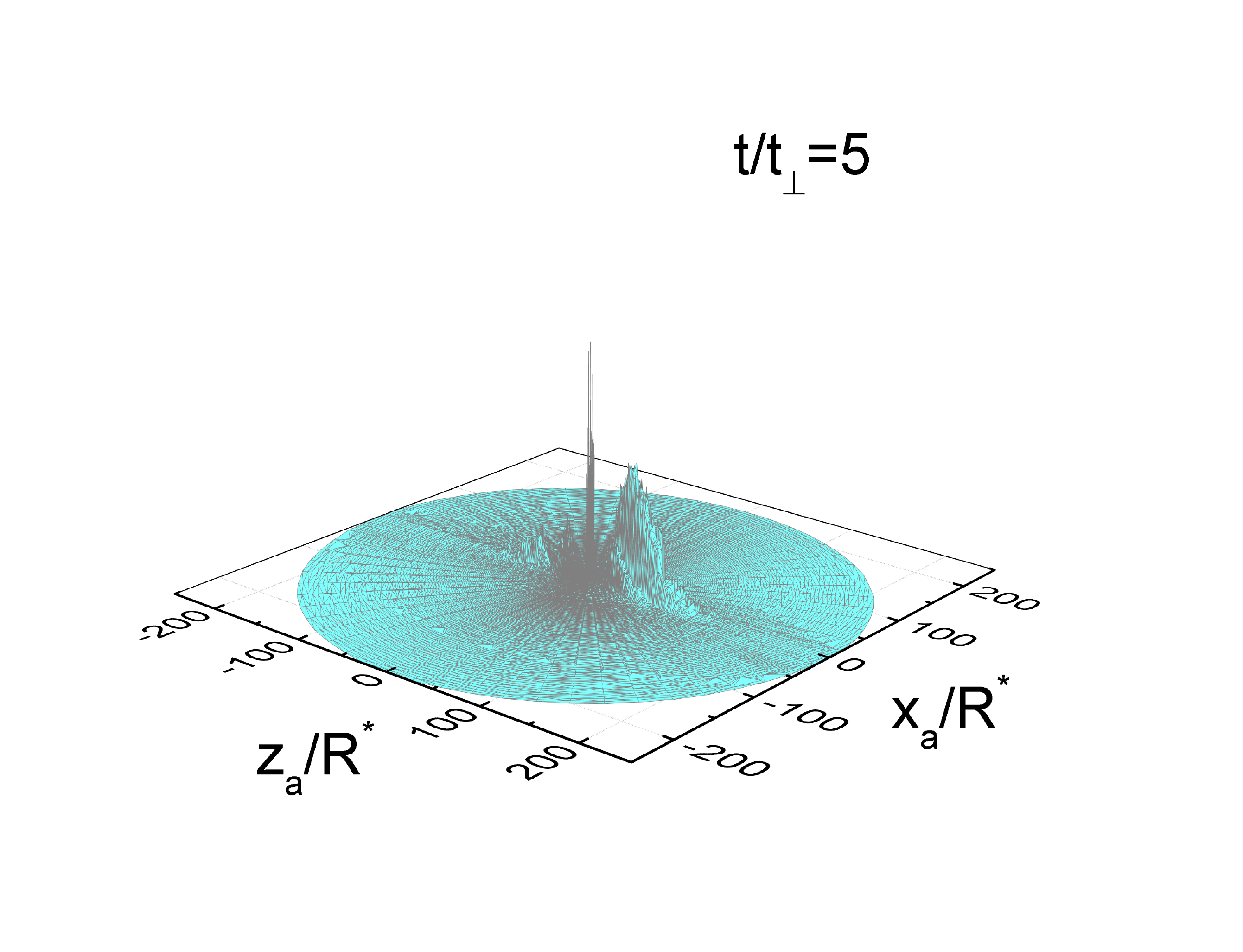}
\vspace{-1.2cm}}
\parbox{0.95\textwidth}{
\includegraphics[width=0.45\textwidth,height=0.45\textwidth]{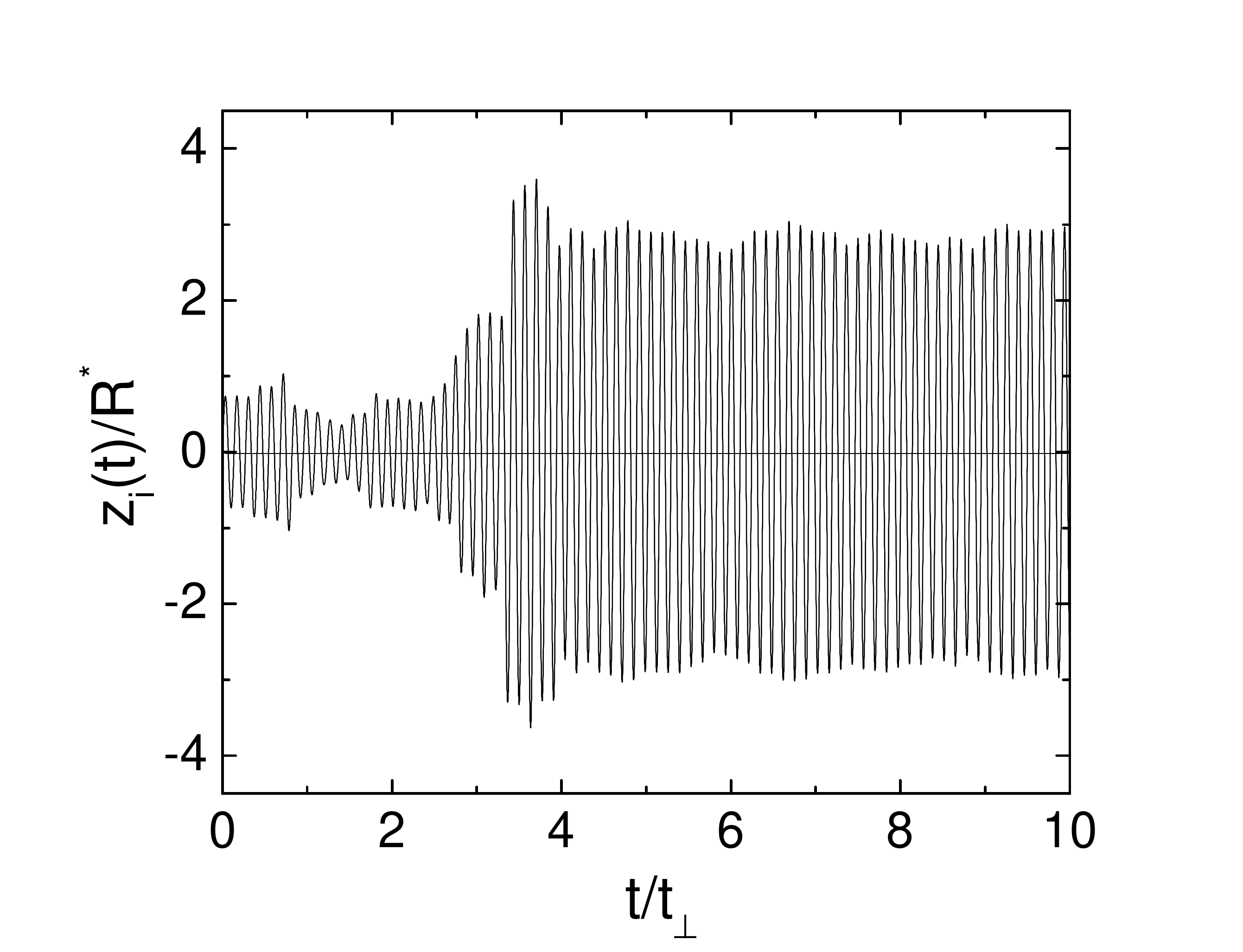}
\includegraphics[width=0.40\textwidth,height=0.4\textwidth]{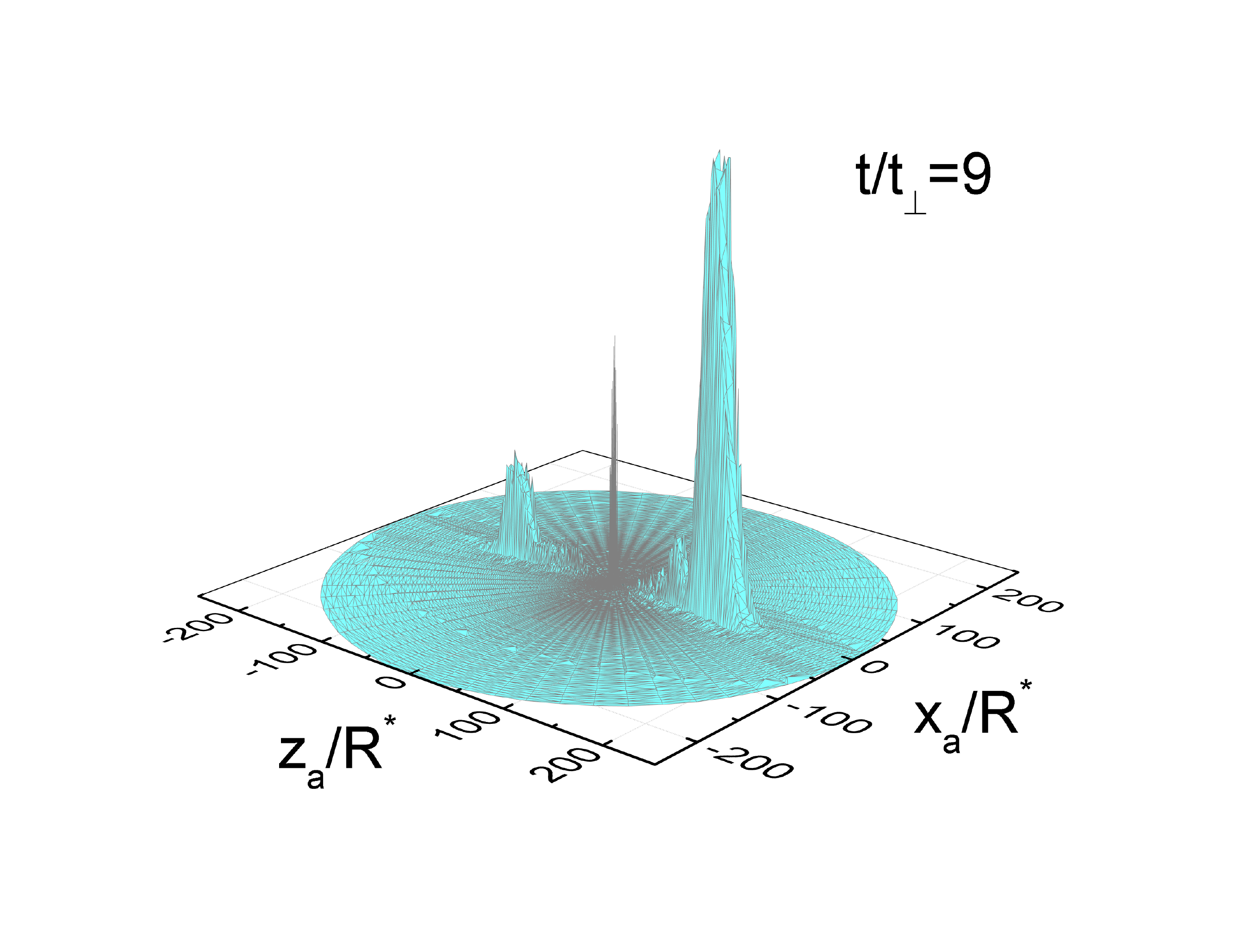}
\vspace{0.2cm}}
\caption{ (color online) The calculated evolution in time of the ion trajectory (left column), being initially at the state with $E_{\perp}=E_{\|}=0.25 E^*=4.25\mu$K, and the atom probability distribution $|r_a\Psi(\vek{r}_a,t)|^2$ (right column) for the atom-ion attractive interaction giving ratio $a_{\perp}/a_s=-3.88$ and for $\omega_{\perp}/\omega^*=0.02$.
}
\label{fig:Fig1}
\end{figure*}
We have performed simulations for different values of the atom-ion interaction $V_{ai}(r(t))$ by varying it from strong repulsion ($a_s \gg 0$) to strong attraction ($a_s\ll 0$). Special attention has been focused on the region near the atom-ion CIR. In the zero-energy limit for the atom and under the static approximation for the ion, the ratio $a_{\perp}/a_s$ [$a_{\perp}=\sqrt{\hbar/(m_a\omega_{\perp})}$] approaches the well known value 1.4603~\cite{Olshanii1998,MelezhikPRA16} when $R^*\ll a_\perp$. Here $R^*=\sqrt{2\mu C_4}/\hbar$ is the characteristic length scale of the atom-ion interaction~(\ref{eq:Vai}), where $\mu$ denotes the atom-ion reduced mass.
The determination of the forward scattering amplitude is performed as follows. For a chosen pair of parameters $b$ and $c$ of the regularised atom-ion potential~(\ref{eq:Vaireg}), we calculate the corresponding scattering length $a_s$ in free space ($\omega_\perp = 0$), which can be easily assessed because of the separation of  the center-of-mass and angular part of the atom-ion wave-function.
Thereafter, we fix the value of transverse frequency $\omega_\perp$, i.e. $a_\perp$, as well as use the previously determined $b$ and $c$ parameters of the interaction potential~(\ref{eq:Vaireg}).
Thus, we simulate the time evolution of the atomic wave-packet and the ion trajectory by integrating simultaneously the equations~(\ref{eq:SE}) and~(\ref{eq:Hamilton}) with the initial conditions~(\ref{eq:initialion}) and~(\ref{eq:initialatom}). Numerical integration has been performed in the time domain from $t=0$ to $t=10\, t_\perp$ with $t_\perp = 2 \pi/\omega_\perp$ defined by the lowest frequency of the problem $\omega_{\perp}\ll \omega_i, \Omega_{rf}$.

An example of such an analysis is illustrated in Fig.~\ref{fig:Fig1}, where the ion coordinates as a function of time are shown together with the atomic probability density distribution $\vert r_a\Psi(\vek{r}_a,t)\vert^2$ at three different times: Before the collision ($t=0$), at the region of the atom-ion interaction ($t=5 t_\perp$), and after the collision ($t=9 t_\perp$, i.e. $t \rightarrow +\infty$). The initial conditions and parameters of the atomic trap, that is, $z_0=70\,R^*$, $a_z=30\,R^*$ and $\omega_{\perp}/\omega^*=0.02$, as well as the atom-ion interaction with $a_{\perp}/a_s=-3.88$, were chosen in such a way that the atom at $t=0$ does not interact with the ion [$V_{ai}(z_a=z_0,r_i) \simeq -C_4/(70R^*)^4\rightarrow 0$]. Here $\omega^* = 2E^*/\hbar$ and $E^*=\hbar^2/[2\mu (R^*)^2]$.  In this case the ion is performing stable oscillations in the Paul trap with initial conditions~(\ref{eq:initialion}) and $E_{\perp}=E_{\|}=0.25\,E^*=4.25 k_B$ $\mu$K. Here, for the atom-ion pair $^6$Li/$^{174}$Yb$^+$ we have: $R^* \simeq 69.77$ nm, that is, $z_0\simeq 4.88\mu$m, while $\omega^* =  2\pi\times 357.16$ kHz, thus $\omega_\perp = 2\pi\times 7.14$ kHz and $t_\perp = 1$ ms. As it is shown in Fig.~\ref{fig:Fig1}, at time $t\sim t_\perp$ the ion begins to experience the vicinity of the atom. This is clearly displayed in the $y_i$-component of the ion motion, where from the initial zero value suddenly large oscillations appear. This is due to the fact the ion has been displaced from the trap centre, and therefore it experiences the radiofrequency fields. Similarly, there is an enhancement of the amplitude of the oscillations of the $x_i$- and $z_i$-components of the ion trajectory. The collision occurs approximately within the time window $t \simeq t_\perp$ up to  $t\simeq 5\,t_\perp$. Afterwards, when the atom and ion leave the range of the atom-ion interaction, they approach the asymptotic region, that is, the two particle do not interact. The ion coordinates reach a steady-state solution, namely the amplitude of the oscillations is approximately constant, but essentially they exceeded their initial values. This indicates that the ion has ``heated up''. Such a finding has been also identified in the master equation approach~\cite{KrychPRA15}, and in Monte Carlo simulations for a trapped ion interacting with classical buffer gas~\cite{DeVoe2009}. On the other hand, the atomic wave-packet splits up in two parts moving forward and backward. Note also the peak in the atomic density distribution remains near to the origin, that is, at the centre of the Paul trap. This indicates that some part of the initial atomic wave-packet is lost or, in other words, an ionic molecule is formed in the Paul trap due to atom-ion collision, which is also a consequence of the negative ratio $a_{\perp}/a_s=-3.88$, i.e. attracting atom-ion interaction. Our estimate gives the following value $P_{mol} \simeq 0.14$ for the probability of creation of molecular ions in such confined atom-ion collisions.
Such a phenomenology resembles the situation of resonant molecule formation in atomic confining traps suggested in Ref.~\cite{MelSch}, where the excess energy is transferred in an excitation of the center-of-mass of the formed molecule. Note that the two-body bound state has been not considered in the study~\cite{KrychPRA15}, where the Born and Markov approximations have been performed.

\begin{figure*}
\parbox{0.95\textwidth}{
\includegraphics[width=0.3\textwidth,height=0.3\textwidth]{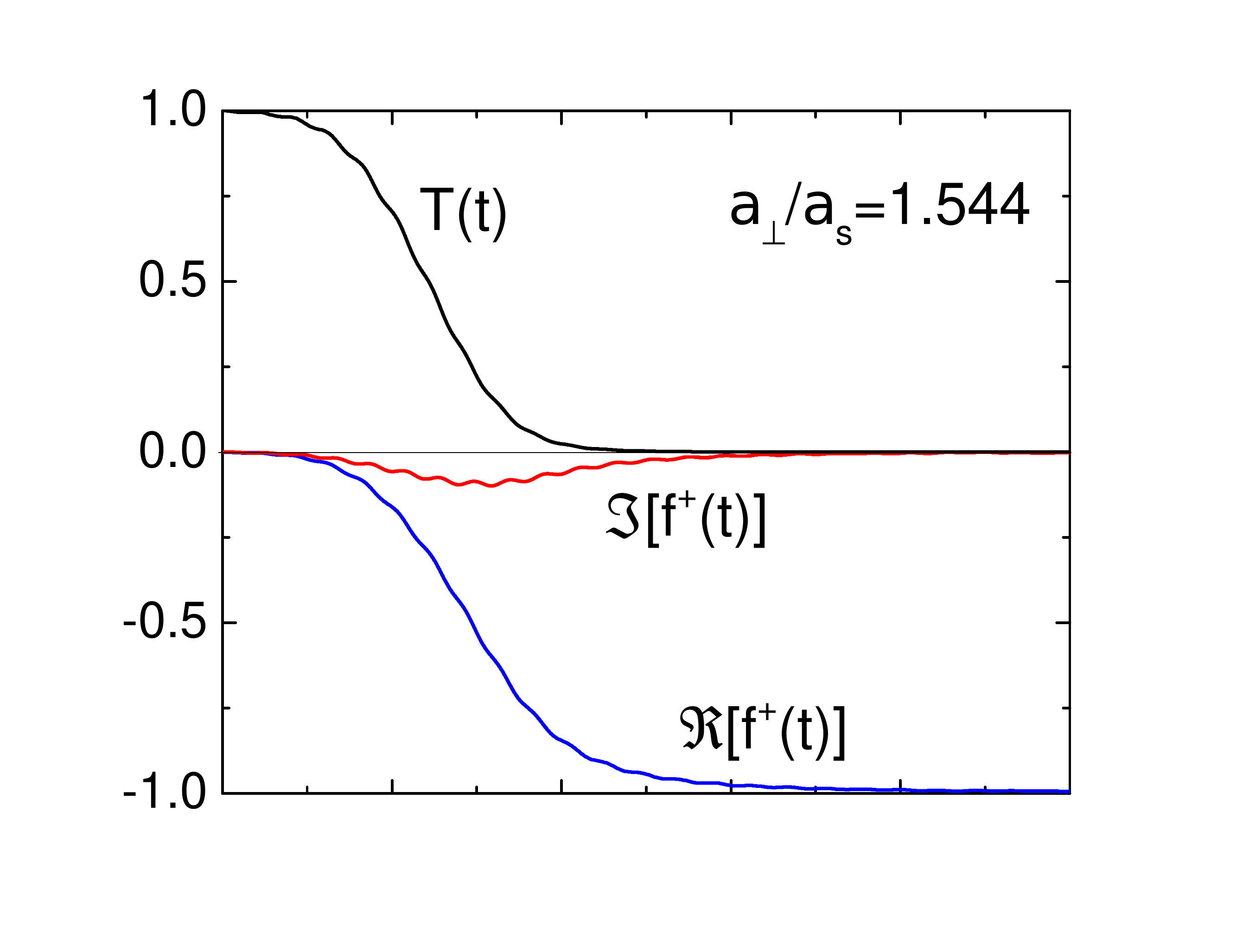}
\includegraphics[width=0.3\textwidth,height=0.3\textwidth]{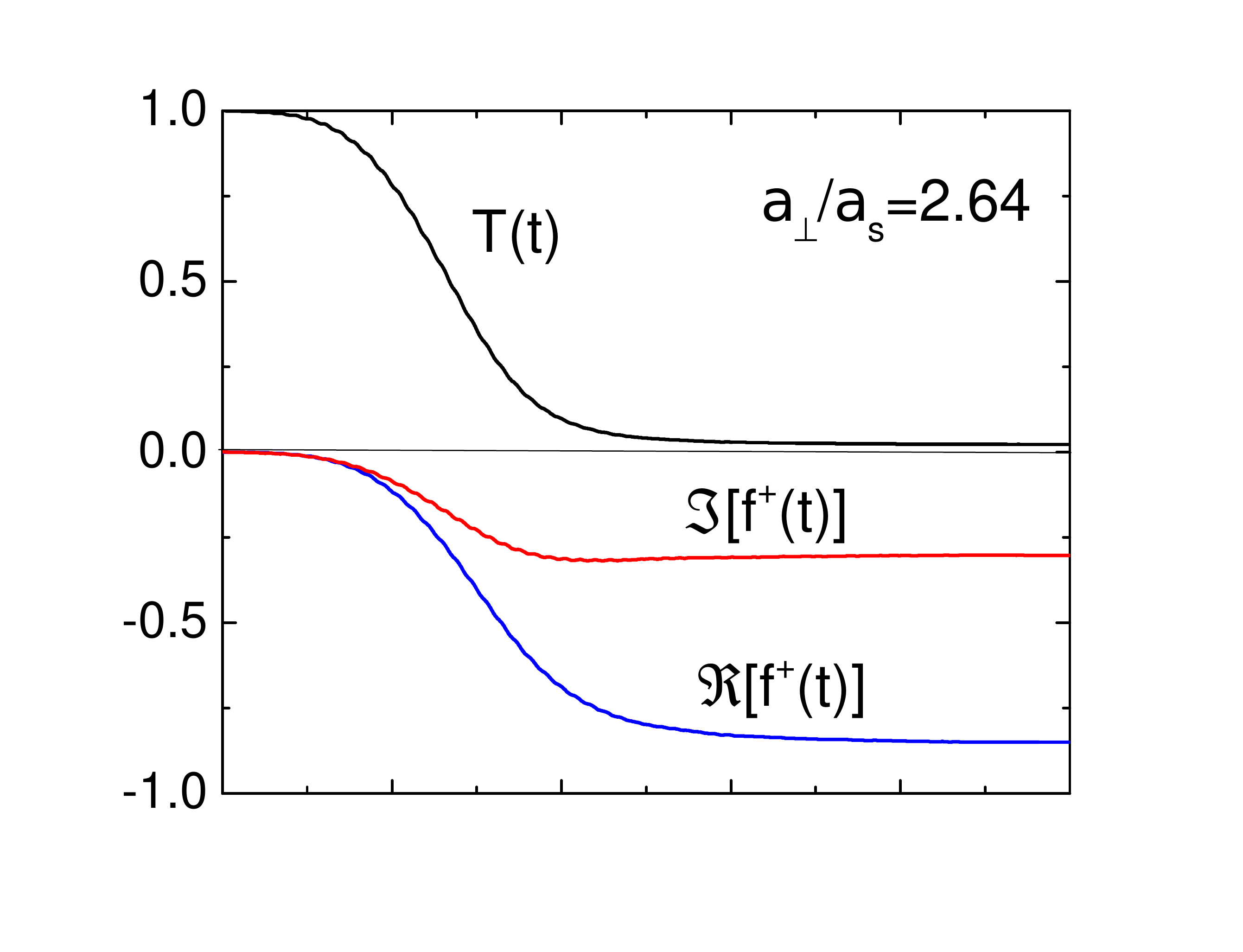}
\includegraphics[width=0.3\textwidth,height=0.3\textwidth]{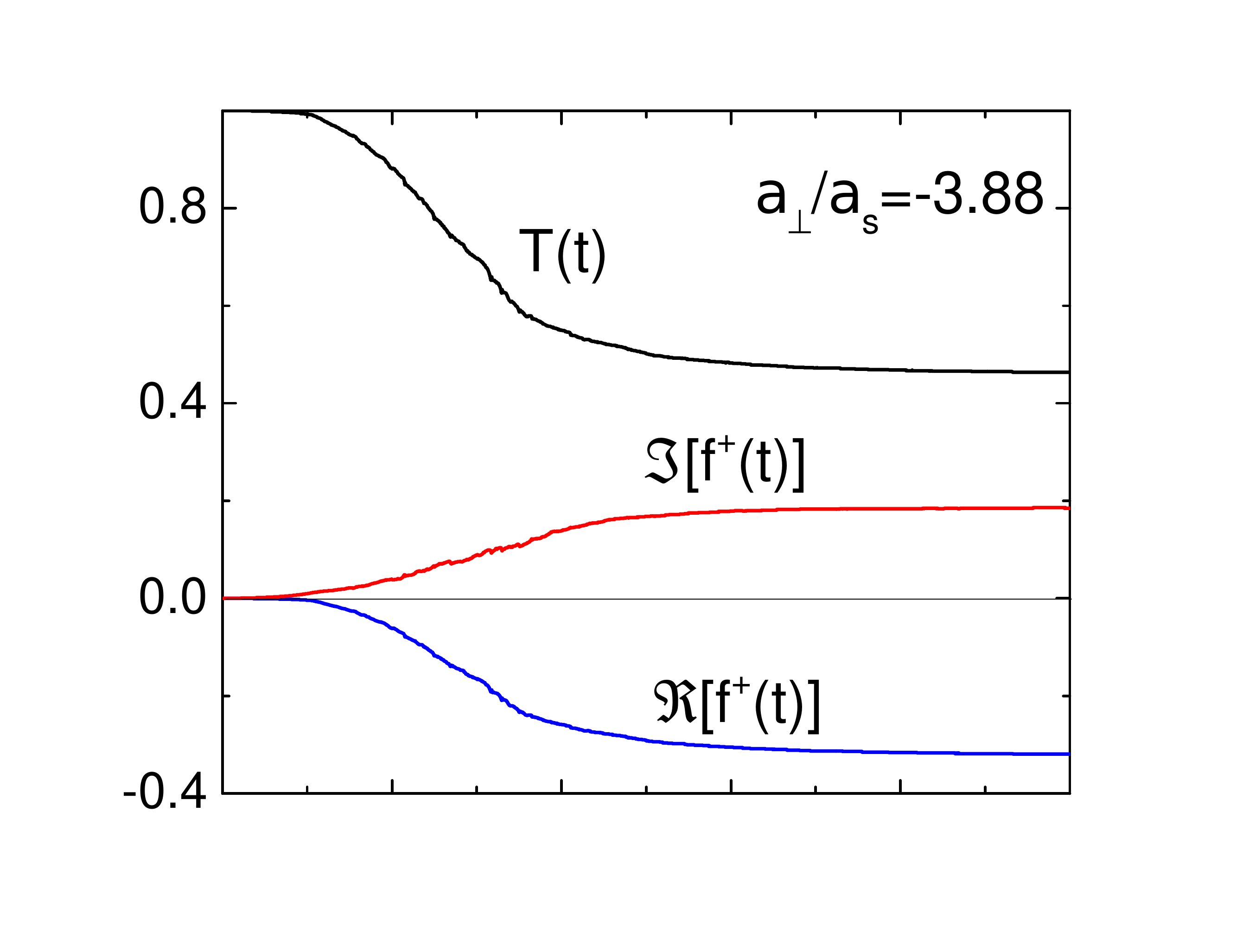}
\vspace{-1.2cm}}
\parbox{0.95\textwidth}{
\includegraphics[width=0.3\textwidth,height=0.3\textwidth]{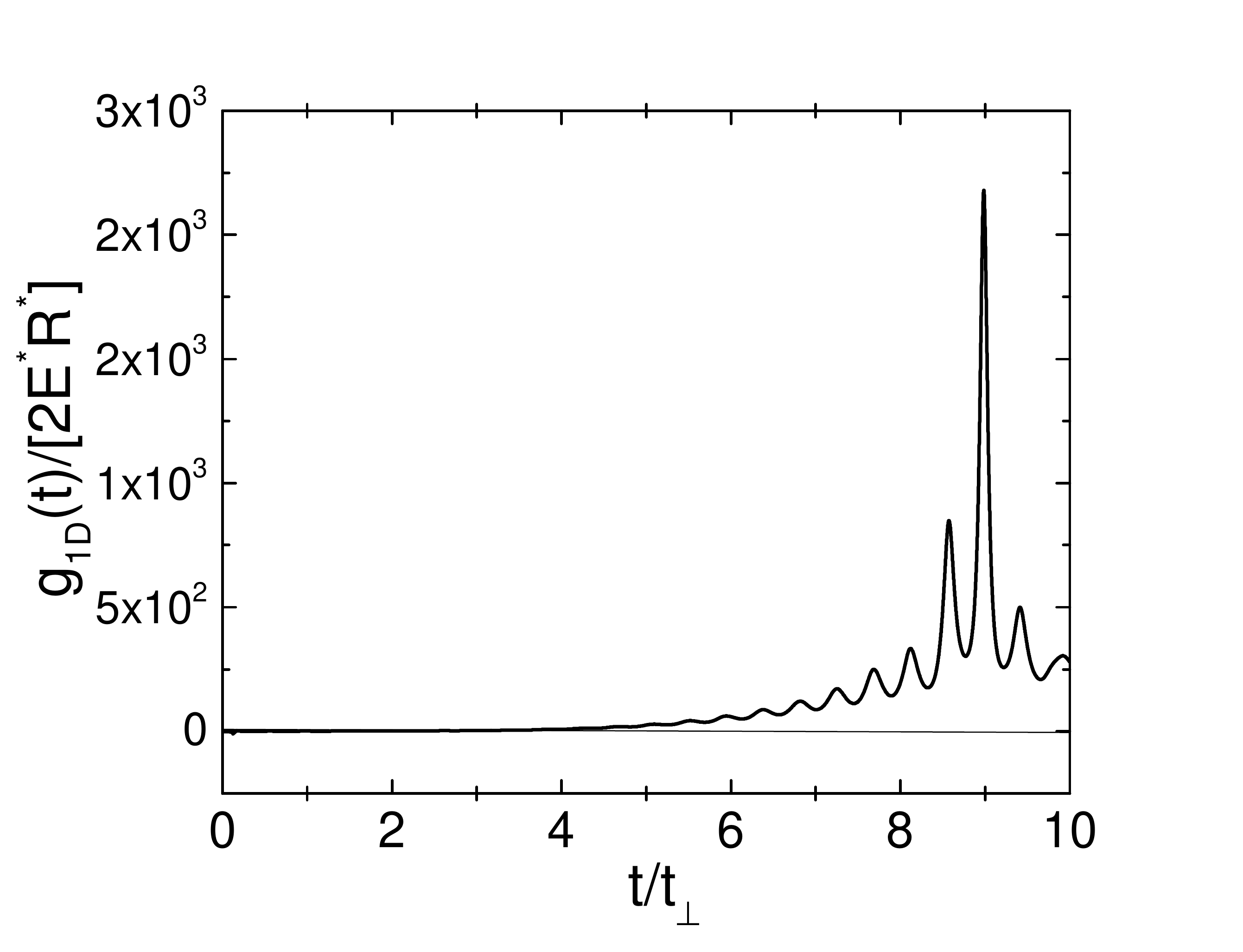}
\includegraphics[width=0.3\textwidth,height=0.3\textwidth]{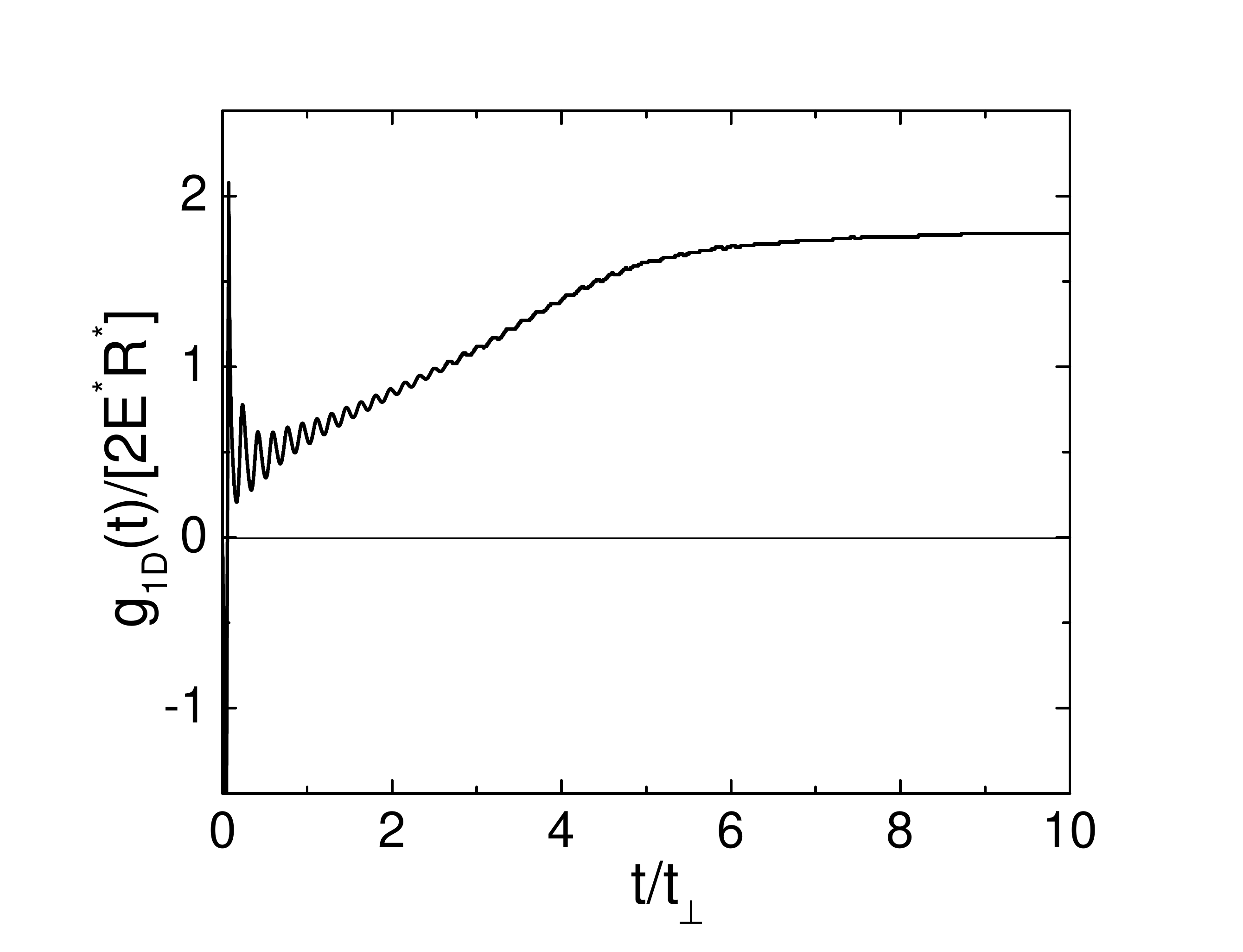}
\includegraphics[width=0.3\textwidth,height=0.3\textwidth]{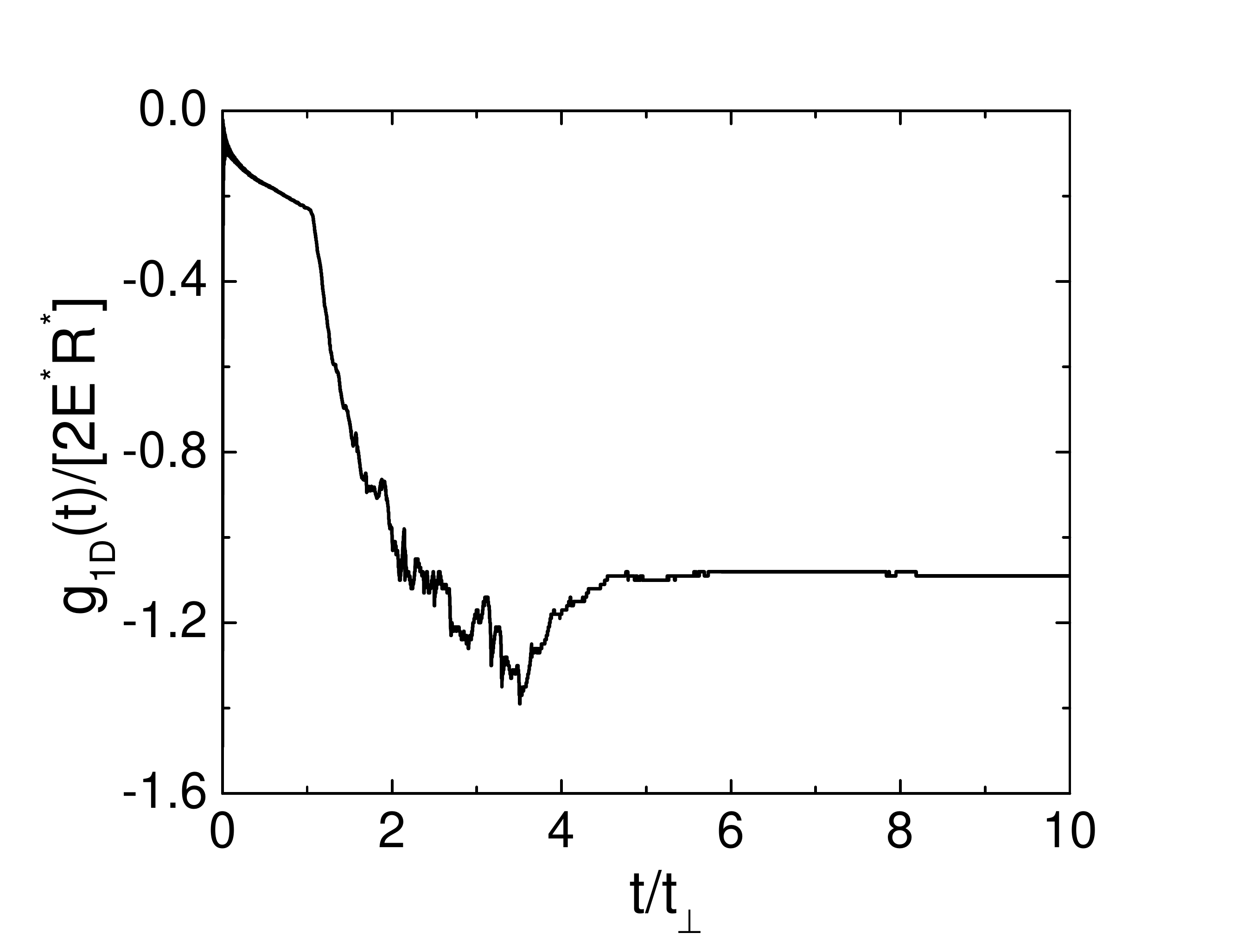}
\vspace{0.2cm}}
\caption{ (color online) The calculated $f^+(t)$, $T(t)$, $g_{1D}(t)$ for the ion being initially at rest (i.e. with zero energy before the collision with atom) for three different values of the ratio $a_{\perp}/a_s$.
}
\label{fig:Fig2}
\end{figure*}

\begin{figure*}
\parbox{0.95\textwidth}{
\includegraphics[width=0.3\textwidth,height=0.3\textwidth]{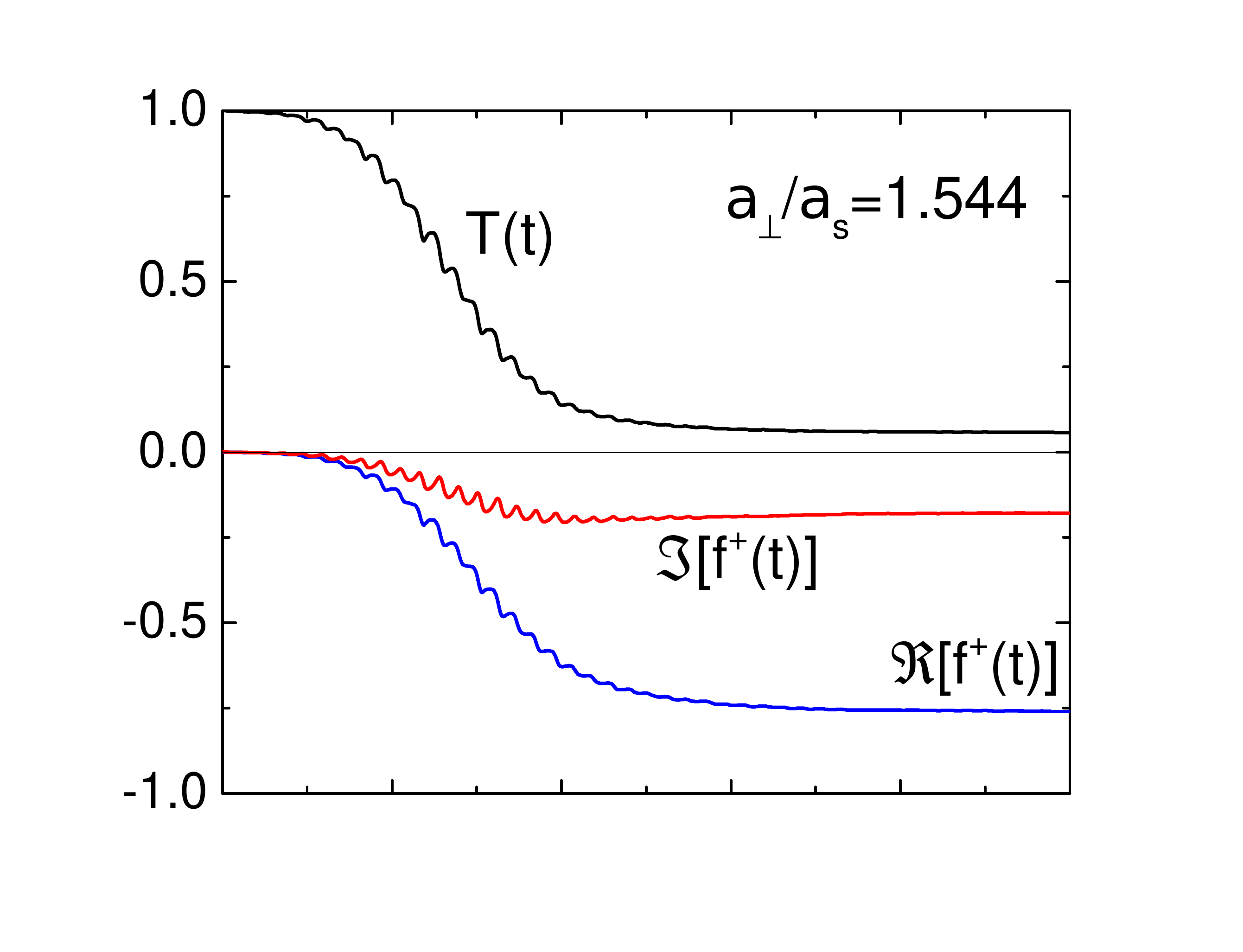}
\includegraphics[width=0.3\textwidth,height=0.3\textwidth]{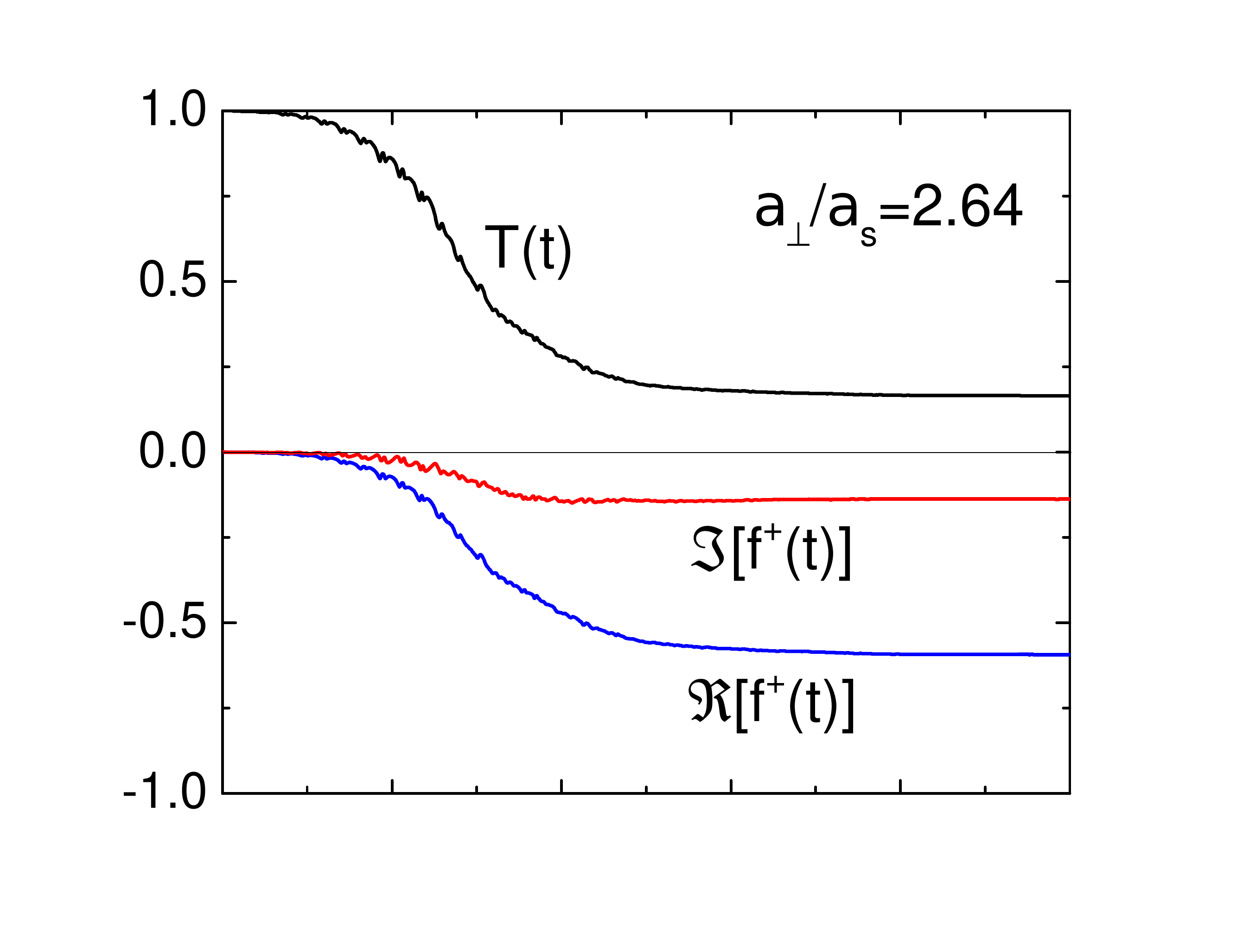}
\includegraphics[width=0.3\textwidth,height=0.3\textwidth]{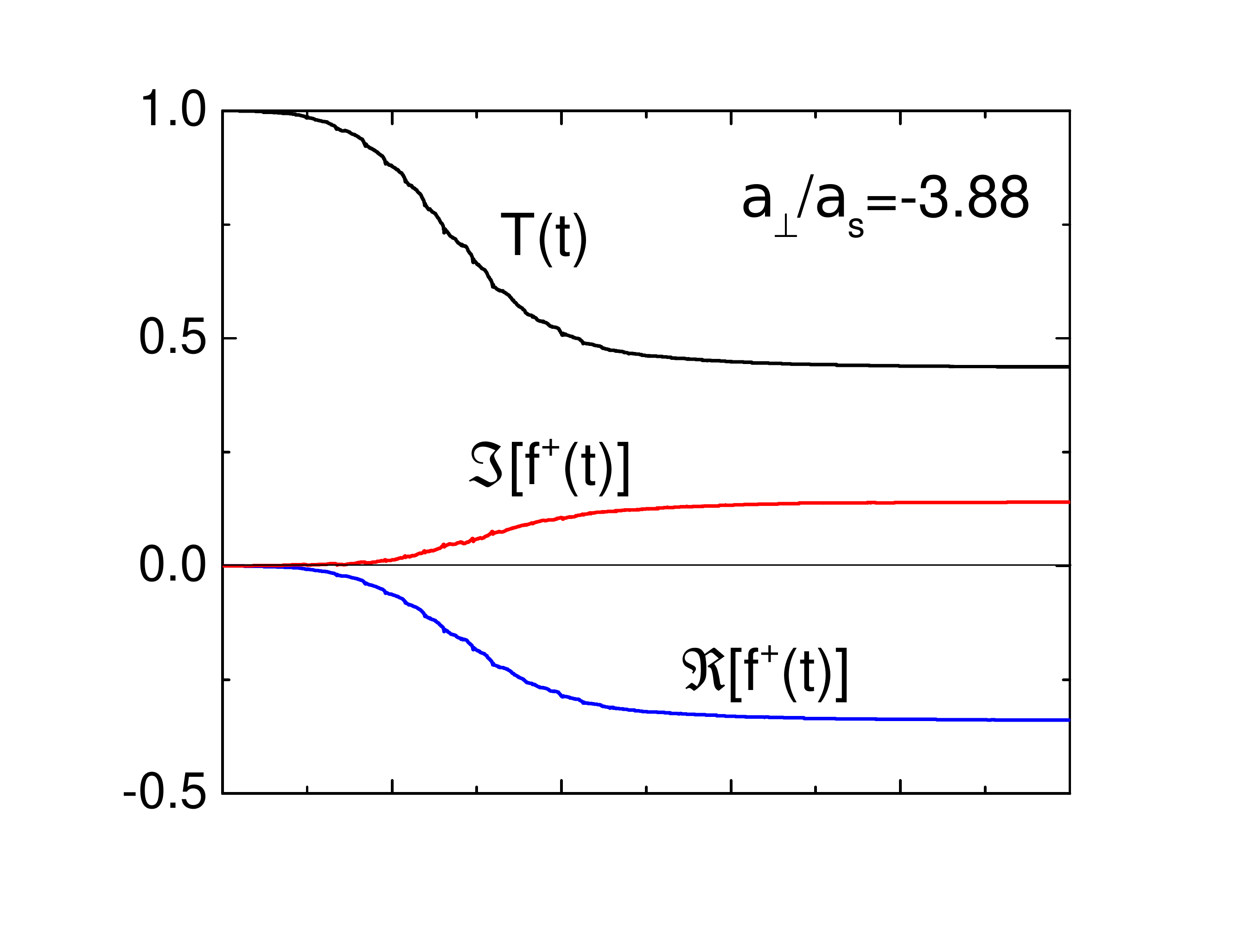}
\vspace{-1.2cm}}
\parbox{0.95\textwidth}{
\includegraphics[width=0.3\textwidth,height=0.3\textwidth]{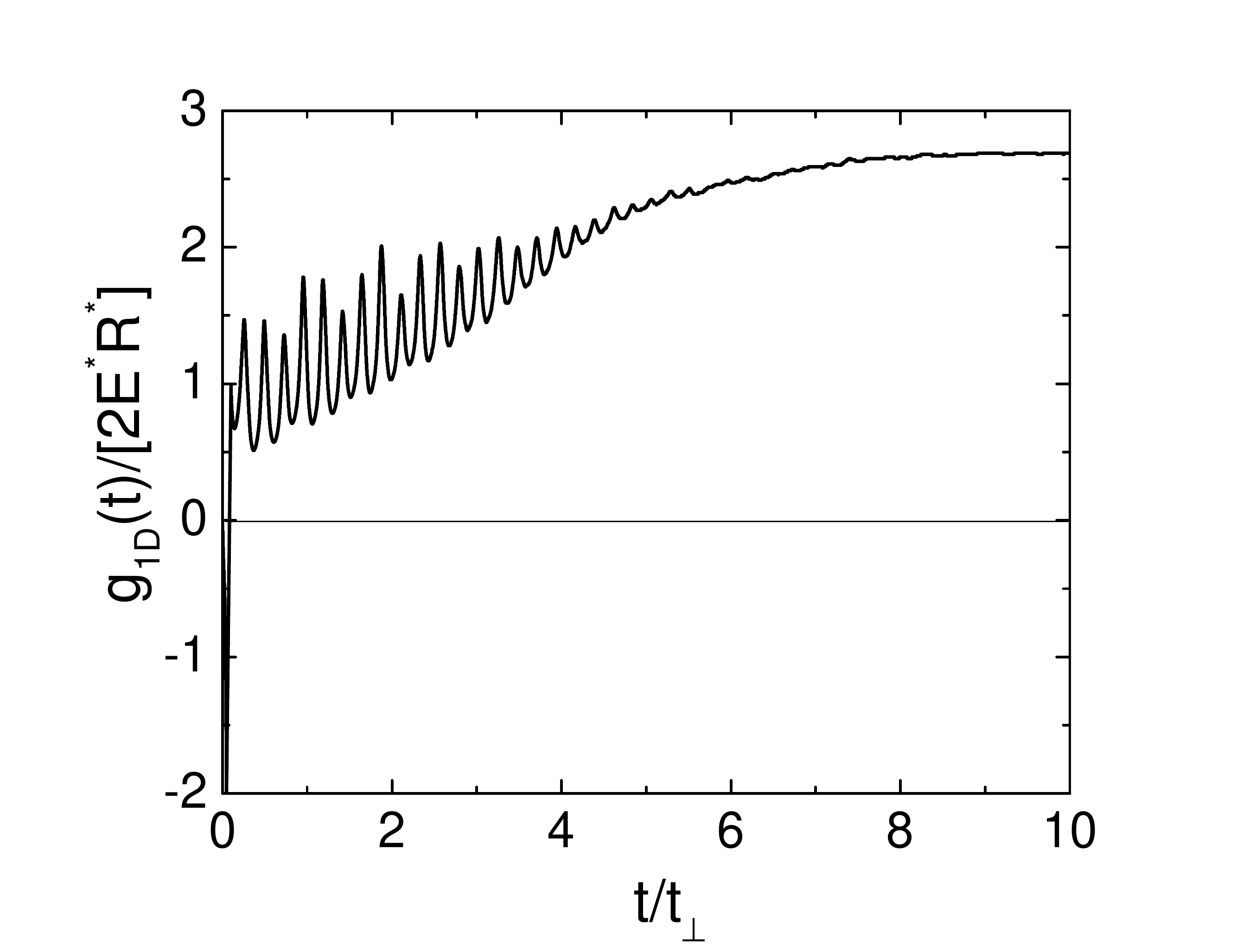}
\includegraphics[width=0.3\textwidth,height=0.3\textwidth]{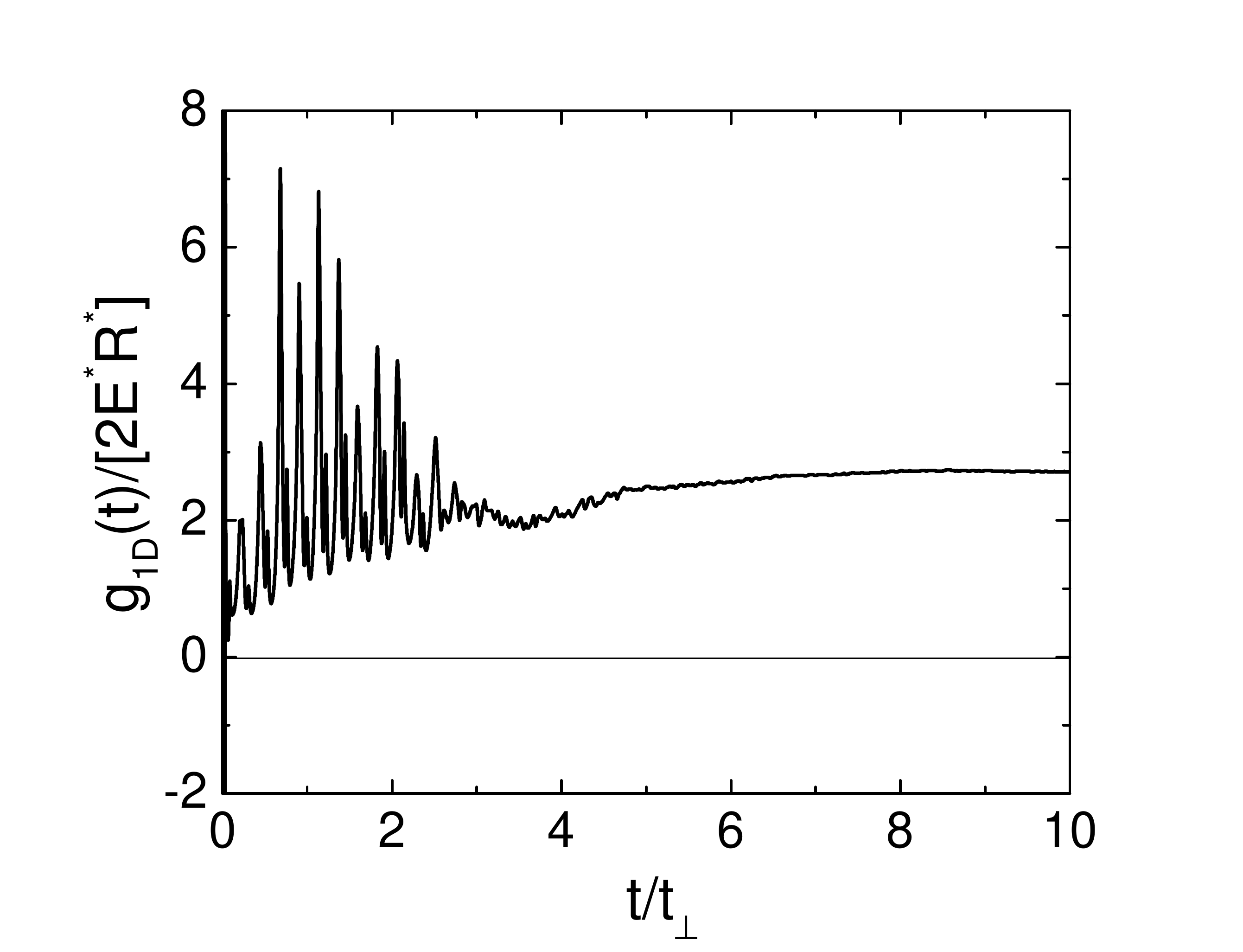}
\includegraphics[width=0.3\textwidth,height=0.3\textwidth]{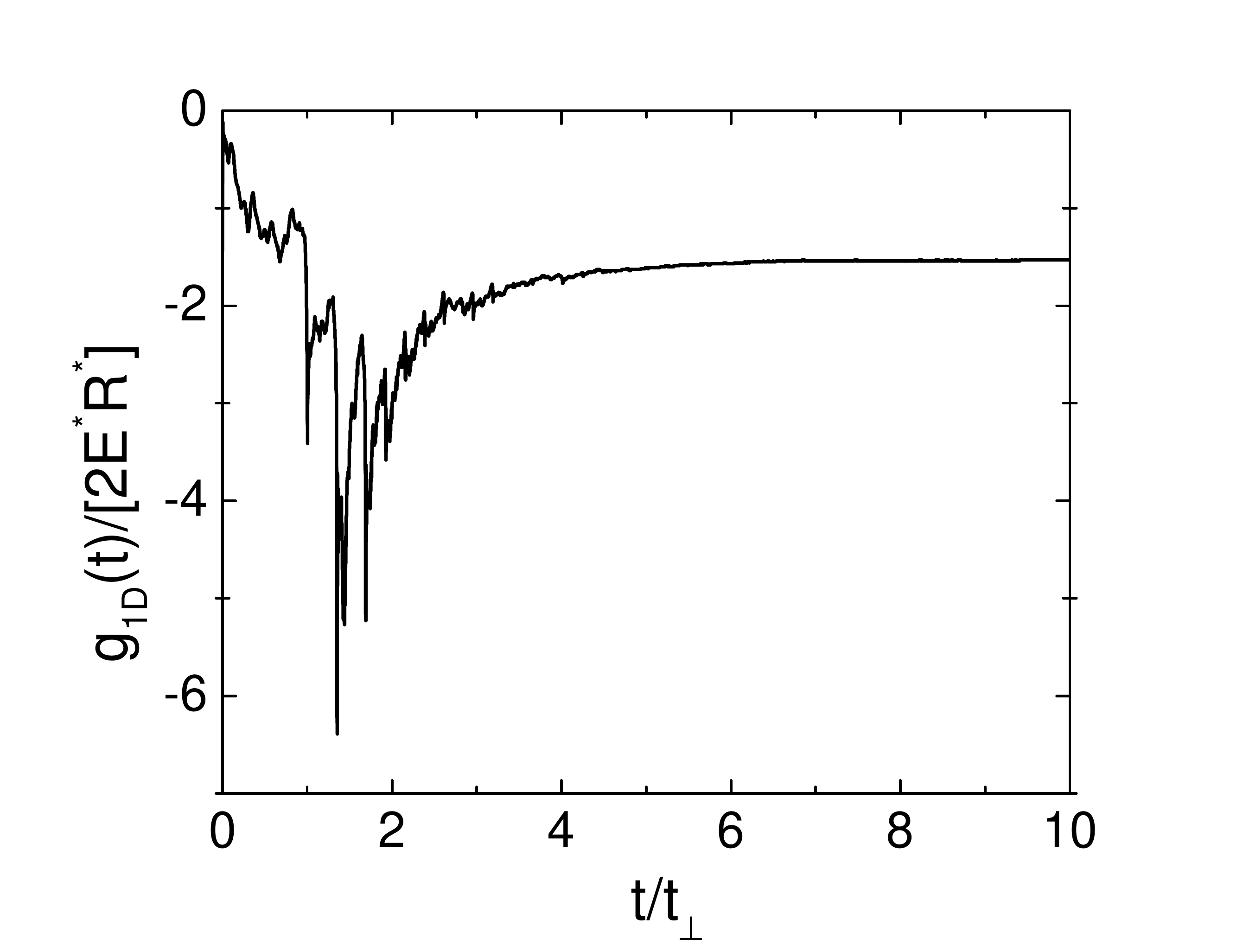}
\vspace{0.2cm}}
\caption{ (color online) The calculated $f^+(t)$, $T(t)$, $g_{1D}(t)$ for an initial ion energy $E_{\perp}=E_{\parallel}=0.25E^*=4.25 k_B\times\mu$K ($p_{\perp}=p_x$) for three different values of the ratio $a_{\perp}/a_s$.
}
\label{fig:Fig3}
\end{figure*}

Thereafter, we have extracted the scattering parameters $f^+(k)$, $T$, and $g_{1D}$ as outlined in Sec. II.D. Figures~\ref{fig:Fig2} and~\ref{fig:Fig3} illustrate the result for two ion energies as well as for three pairs $(b,c)$ of the atom-ion potential~(\ref{eq:Vaireg}): $a_{\perp}/a_s=1.544$, which corresponds to the case of the atom-ion CIR obtained in the static ion approximation~\cite{MelezhikPRA16}; $a_{\perp}/a_s=2.64$, i.e. a rather strong repulsion between atom and ion; $a_{\perp}/a_s=-3.88$, that is, considerable attraction between the atom and the ion. In Fig.~\ref{fig:Fig2} the calculated scattering parameters $f^+(k)$, $T$, and $g_{1D}$ are presented for the case of the ion being initially ($t=0$) at rest in the centre of the Paul trap with zero initial energy ($E_{\perp}=E_{\parallel}=0$). These results demonstrate the efficiency of the computational procedure outlined in Sec. II.D for extracting the scattering parameters for different strengths of the atom-ion interaction~(\ref{eq:Vaireg}) including strong repulsion and attraction between atom and ion.
We underline that our analysis confirms that if the initial ion energy is zero ($E_{\perp}=E_{\parallel}=0$), we find that the position of the CIR coincides rather well with the result obtained in the static ion approximation~\cite{MelezhikPRA16}. In the plot of the left top panel (i.e. $a_{\perp}/a_s=1.544$) of Fig.~\ref{fig:Fig2} it is shown the behaviour of the scattering parameters, which is characteristic of a CIR, i.e. $\Re [f^{+}]\rightarrow -1$, $\Im [f^{+}]\rightarrow 0$ and $g_{1D} \rightarrow \pm \infty$~\cite{Olshanii1998,MelezhikPRA16}. It also shows that away from the CIR position, the effective coupling constant approaches finite values, positive one for the repulsive atom-ion interaction ($a_{\perp}/a_s=2.64$) and negative one for the attractive atom-ion potential ($a_{\perp}/a_s=-3.88$). Since the result at $a_{\perp}/a_s=1.544$ corresponds to the resonant scattering near the CIR, we observe here the dramatic enhancement of the effective coupling constant $g_{1D}$ with respect to the nonresonant $g_{1D}$ at $a_{\perp}/a_s=2.64$ and $-3.88$. Strong oscillations in the asymptotic region $t\rightarrow +\infty$ observed in $g_{1D}$ at $a_{\perp}/a_s=1.544$ with the time period $\sim 2\pi/(2 \omega_{\perp})$ corresponds to virtual transitions between the entrance channel and the first closed excited state, which separates by the energy threshold by the amount $2\hbar\omega_{\perp}$. This is perfectly consistent with the physical interpretation of the CIR as a resonance in the first closed transverse channel~\cite{Berg}.
Then, we have extended our investigation to the case of the ion oscillating before the collision in the Paul trap with a rather large energy ($E_{\perp}=E_{\parallel}=0.25 E^* = 4.25\mu $K). As it is shown in Fig.~\ref{fig:Fig3}, if we increase the ion energy the position of the CIR is shifted from the point $a_{\perp}/a_s=1.544$, which corresponds to the CIR in the case of the ion at rest before the collision. For these parameters, the atom-ion potential gives a repulsive coupling constant of finite value $g_{1D} /(E^*R^*)=2.7$ (see low left panel in Fig.~\ref{fig:Fig3}). Effective coupling constants $g_{1D}$ calculated for the nonresonant cases $a_{\perp}/a_s =2.64$ and $=-3.88$ are also shifted with increasing ion energy with respect to the case of the ion at rest. We note that the oscillation frequency of the coupling constant $g_{1D}(t)$ in all nonresonant cases presented in Figs.~\ref{fig:Fig2} and \ref{fig:Fig3} is considerably smaller than in the resonant case (CIR) and essentially determined by $\omega_i$. On the other hand, in the case $a_{\perp}/a_s=-3.88$, the oscillations of $g_{1D}(t)$ are determined by $\Omega_{rf}$ of the micromotion.


\begin{figure}
\centering\includegraphics[scale=0.31]{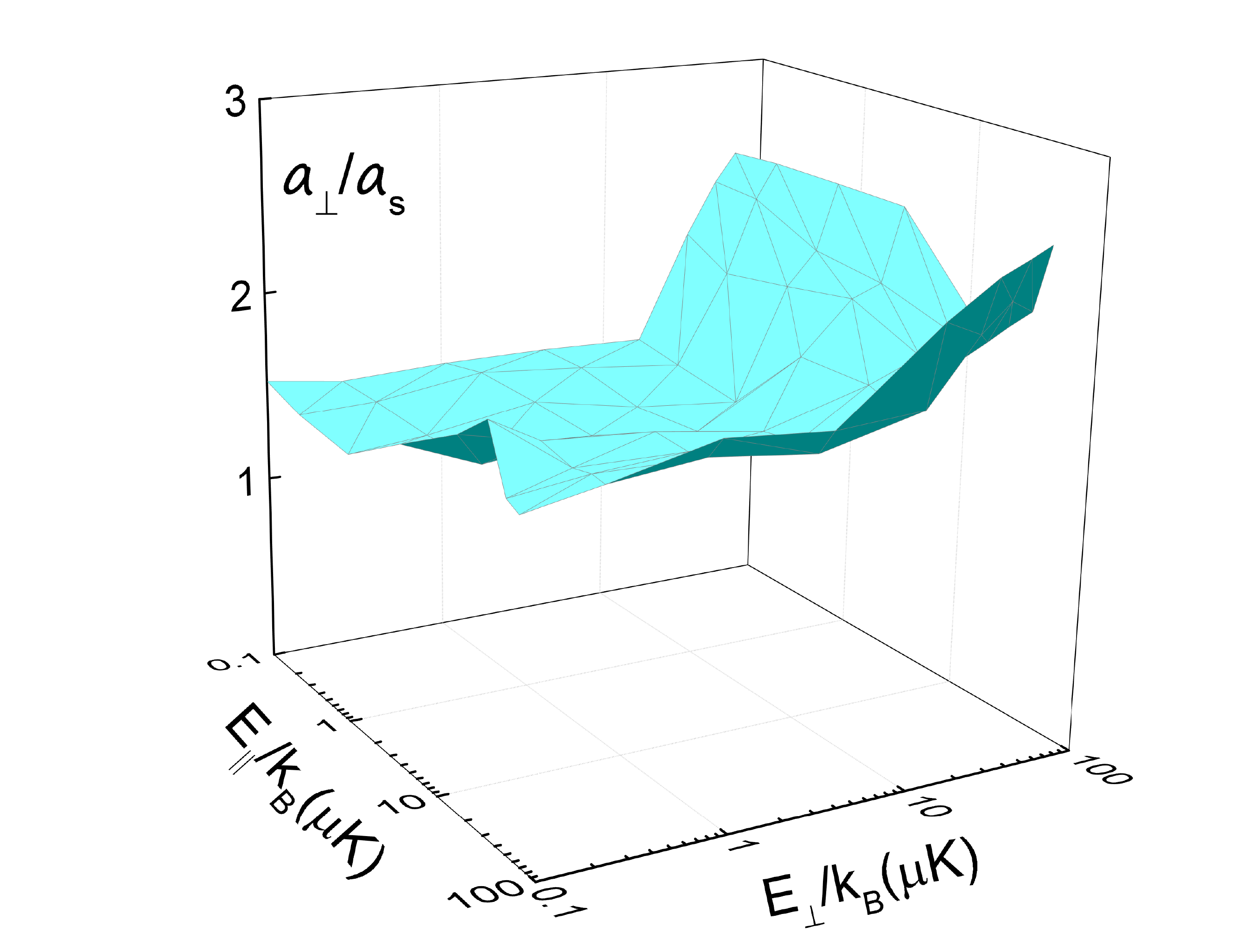}
\caption{ (color online) The calculated position $a_{\perp}/a_s$ of the atom-ion CIR as a function of the transversal $E_{\perp}$ and longitudinal $E_{\parallel}$ kinetic energy of the ion in the secular harmonic trap approximation~(\ref{eq:Usec}) with frequencies ($\omega_{z}=2\pi\times 45$ kHz and $\omega_{x,y}=2\pi\times 150$ kHz).}
\label{fig:Fig4}
\end{figure}

We continue by analysing the dependence of the position of the confinement-induced resonance on the initial (mean) ion energy in the Paul trap. To this aim, we first investigate the scenario for the secular time-independent trap
\begin{align}
\label{eq:Usec}
U_{sec}(\vek{r}_i) &= \frac{m_i}{2}\left[\omega_{xy}^2(x_i^2+y_i^2)+\omega_z^2 z_i^2\right]
\end{align}
with the frequencies $\omega_z=2\pi\times 45$ kHz and $\omega_{x,y}=2\pi\times 150$ kHz.
The positions of the atom-ion CIR were obtained by looking for the positions of the singular points in the coupling constant $g_{1D}(E_{\perp},E_{\|})$~(\ref{eq:g1D}), namely when the coupling constant diverges. In Fig.~\ref{fig:Fig4} we present the calculated dependence of the CIR position $a_{\perp}/a_s$ on the transversal $E_{\perp}$ and longitudinal $E_{\parallel}$ ion energies. The values of $a_{\perp}/a_s$ given in Fig.~\ref{fig:Fig4} at $E_{\perp} \sim E_{\parallel} \rightarrow 0$ confirm the result obtained in Ref.~\cite{MelezhikPRA16} in the static ion approximation in the limit $R^*\ll a_\perp$ for which the atom-ion CIR position coincides with the well-known result $a_{\perp}/a_s=1.4603$~\cite{Olshanii1998}. Moreover, it is found that the calculated position of the atom-ion CIR is fixed quite well near the constant value $a_{\perp}/a_s\simeq 1.5$ in the square domain $E_{\perp}/k_B, E_{\|}/k_B \leq 10\,\mu$K. In other words, in the secular harmonic trap approximation~(\ref{eq:Usec}), the CIR position is stabilised near the value (1.4603) obtained in the static approximation for the ion (independent of the ion mean energy) if the ion transversal and longitudinal initial energies do not exceed the value of $10\,\mu$K (see the almost flat region in Fig.~\ref{fig:Fig4}), which is close to the the s-wave threshold energy $E^*=\hbar^2/[2\mu (R^*)^2] \simeq 6.4\,\mu$K \cite{CetinaPRL12}.

\begin{figure*}
\parbox{0.95\textwidth}{
\includegraphics[width=0.45\textwidth,height=0.4\textwidth]{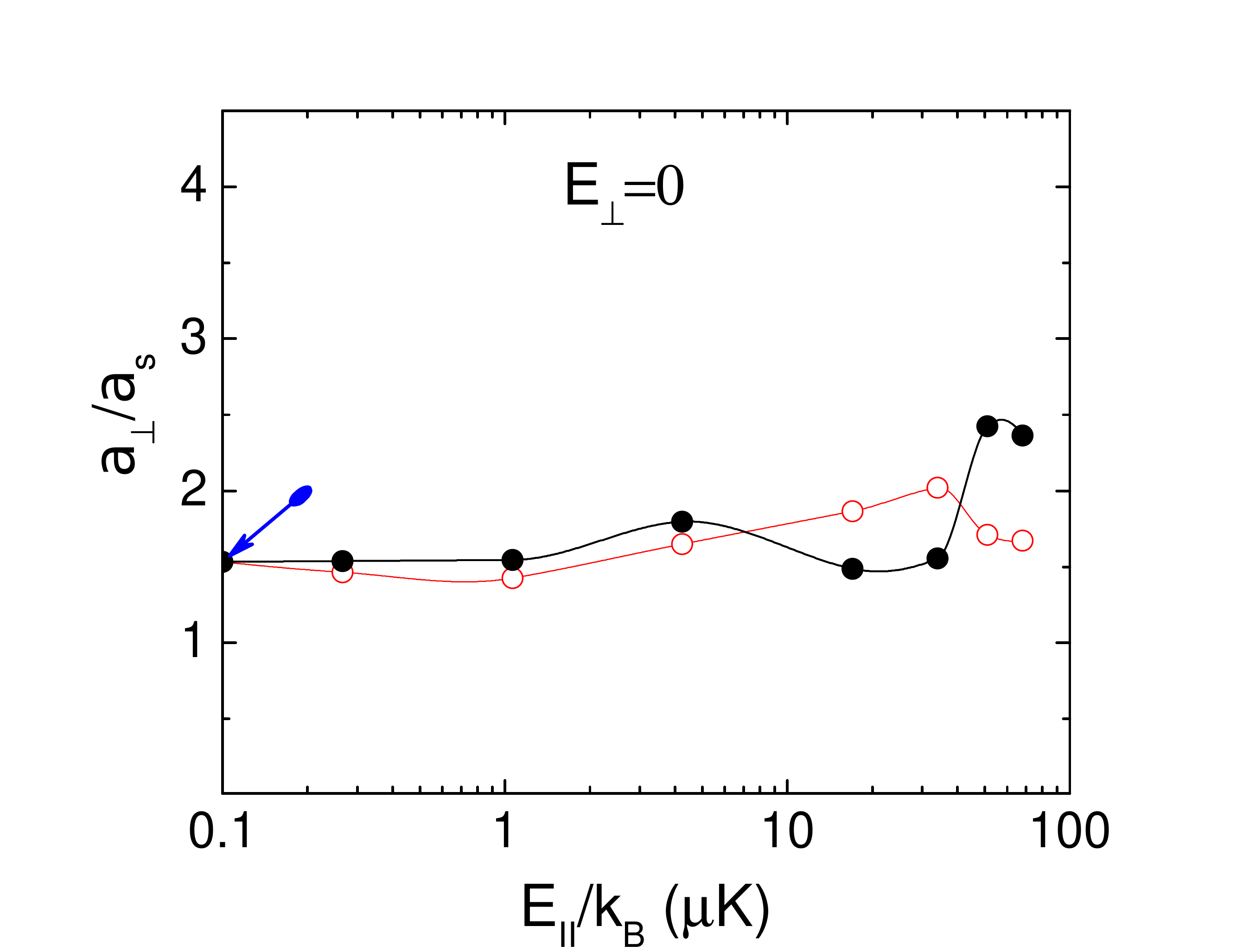}
\includegraphics[width=0.45\textwidth,height=0.4\textwidth]{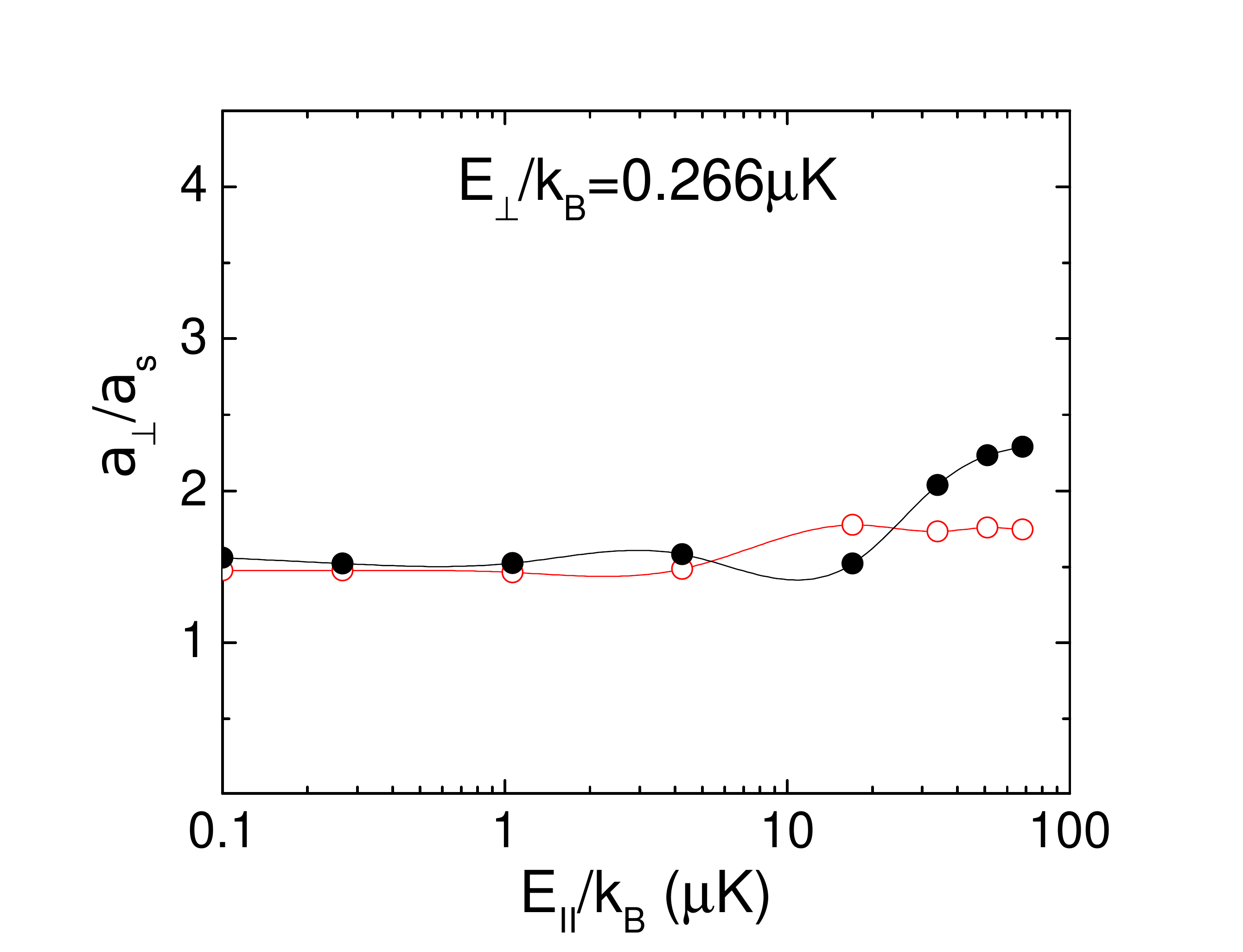}
\vspace{-.5cm}}
\parbox{0.95\textwidth}{
\includegraphics[width=0.45\textwidth,height=0.4\textwidth]{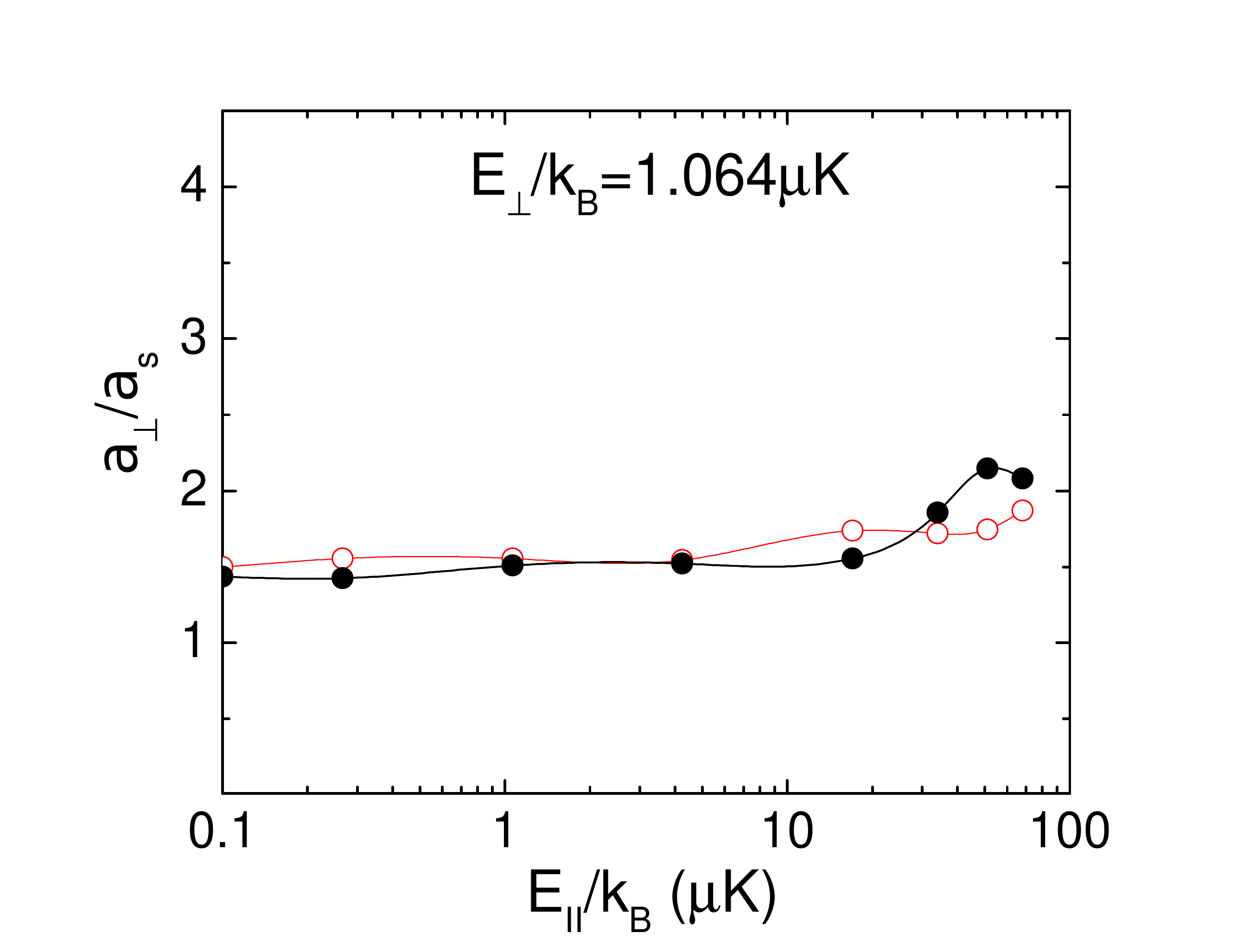}
\includegraphics[width=0.45\textwidth,height=0.4\textwidth]{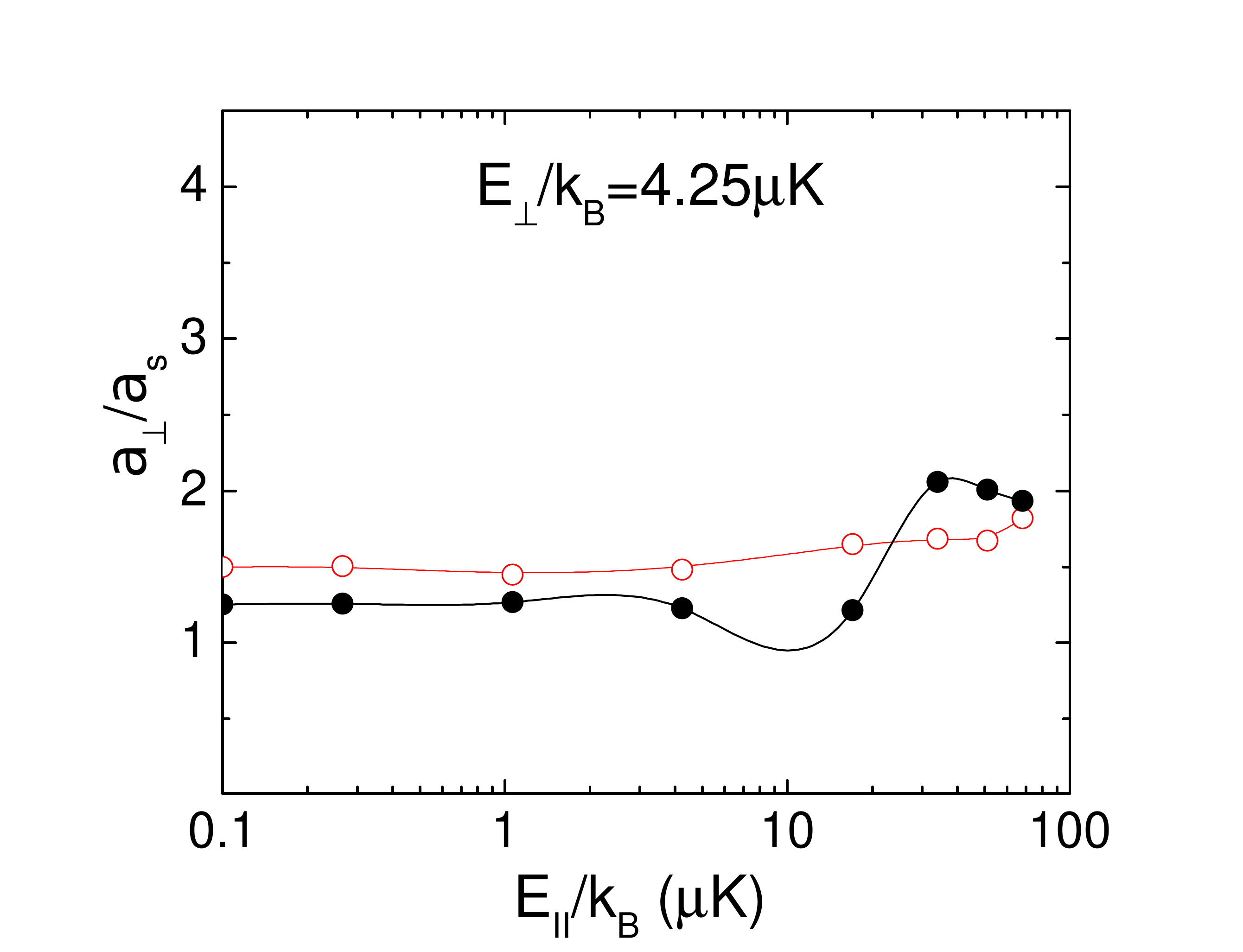}
\vspace{-.5cm}}
\parbox{0.95\textwidth}{
\includegraphics[width=0.45\textwidth,height=0.4\textwidth]{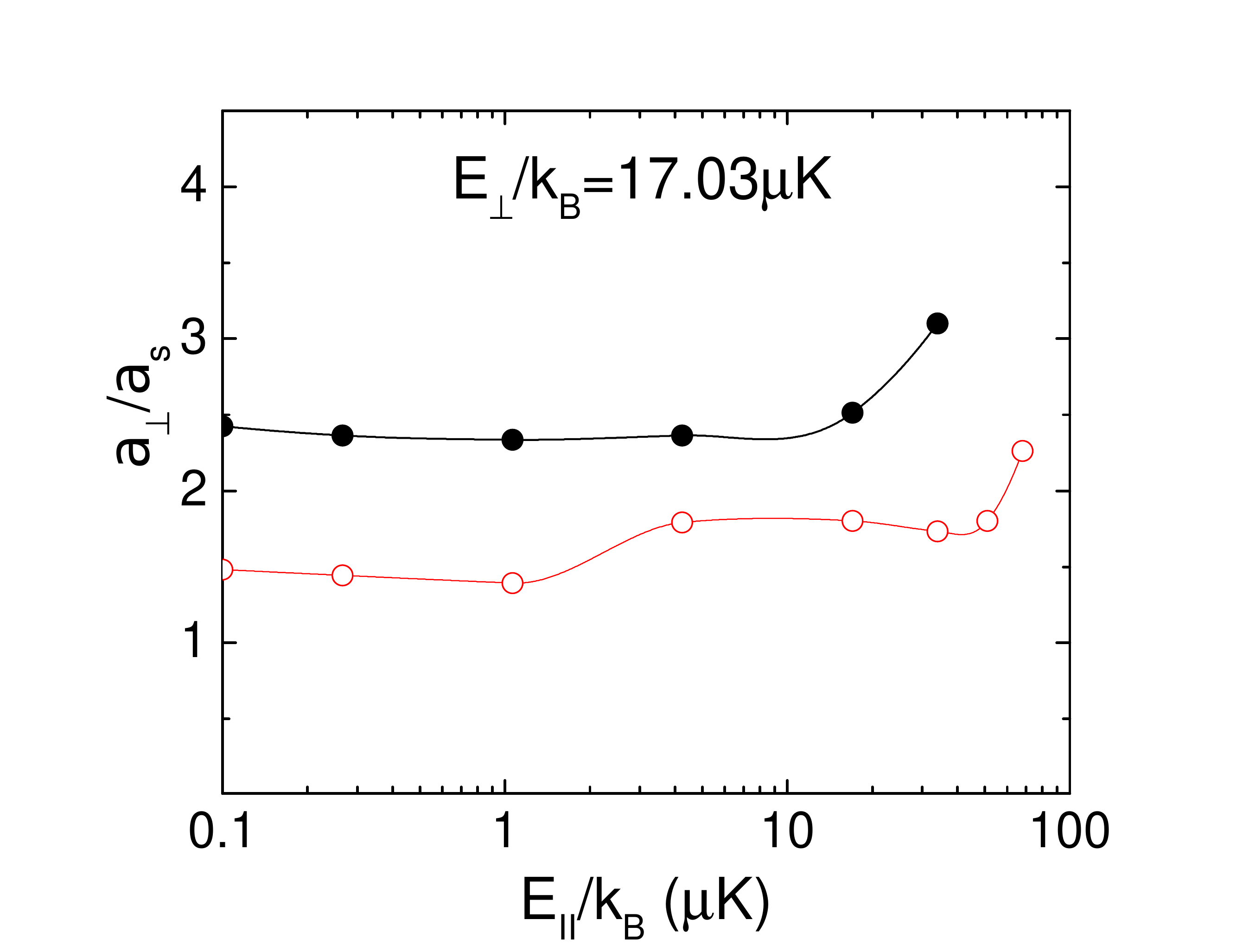}
\includegraphics[width=0.45\textwidth,height=0.4\textwidth]{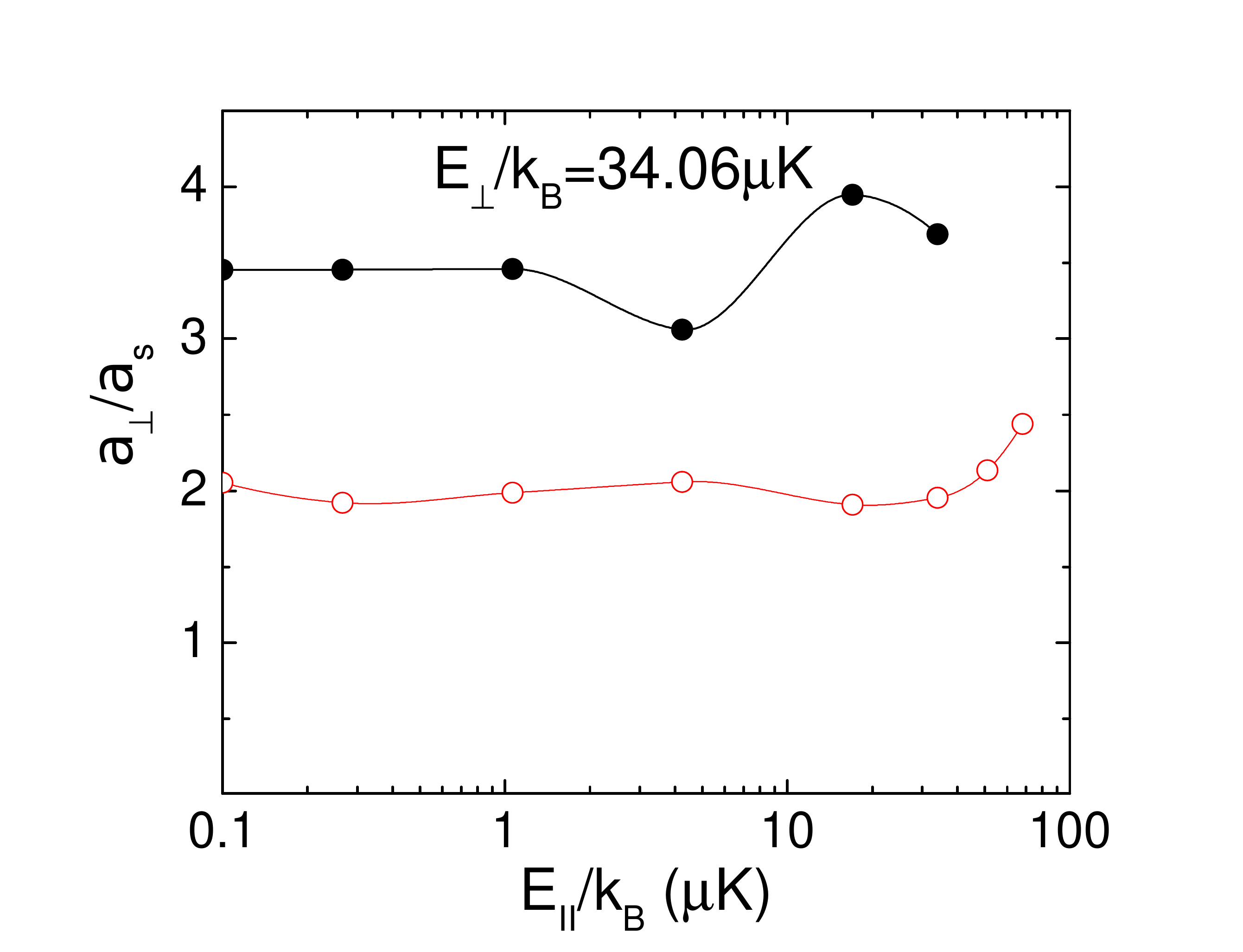}}
\vspace{0.5cm}
\caption{The dependence of the CIR position $a_{\perp}/a_s$ on the transversal $E_{\perp}$ and longitudinal $E_{\parallel}$ ion energies before the collision for the atomic trap with $\omega_{\perp}=0.02\,\omega^* =2\pi\times 7.1 $ kHz. The (blue) arrow in the top left panel (i.e. $E_{\perp}=0$) indicates the atomic CIR position 1.4603 at the zero-energy limit~\cite{Olshanii1998}. Here the results of the calculations with the time-dependent Paul trap~(\ref{eq:Uion}) are given as black cycles connected by (black) solid lines. The results obtained in the secular harmonic trap approximation~(\ref{eq:Usec}) are given as open cycles connected by (red) thin lines.
}
\label{fig:Fig5}
\end{figure*}

\begin{figure*}
\parbox{0.95\textwidth}{
\includegraphics[width=0.45\textwidth,height=0.4\textwidth]{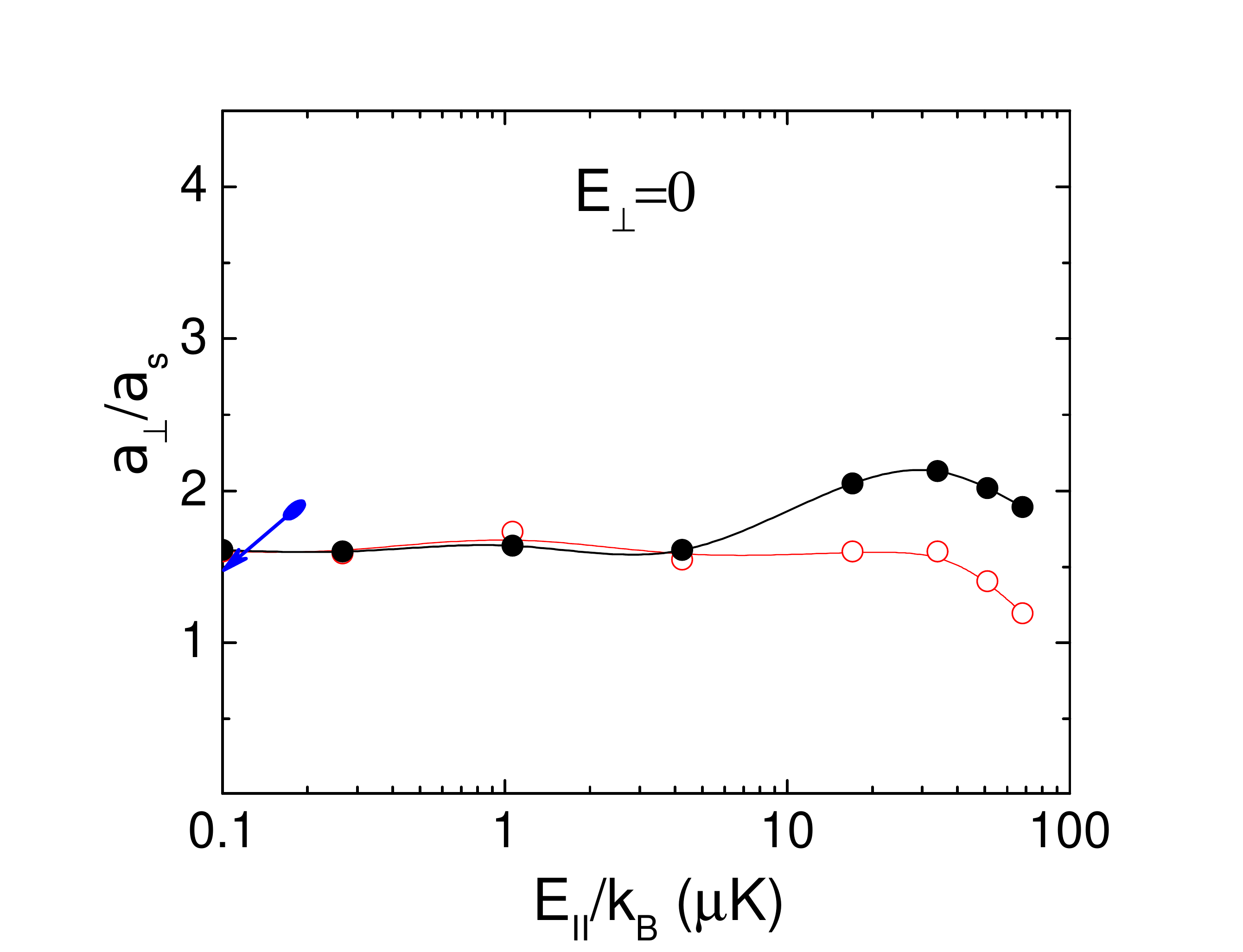}
\includegraphics[width=0.45\textwidth,height=0.4\textwidth]{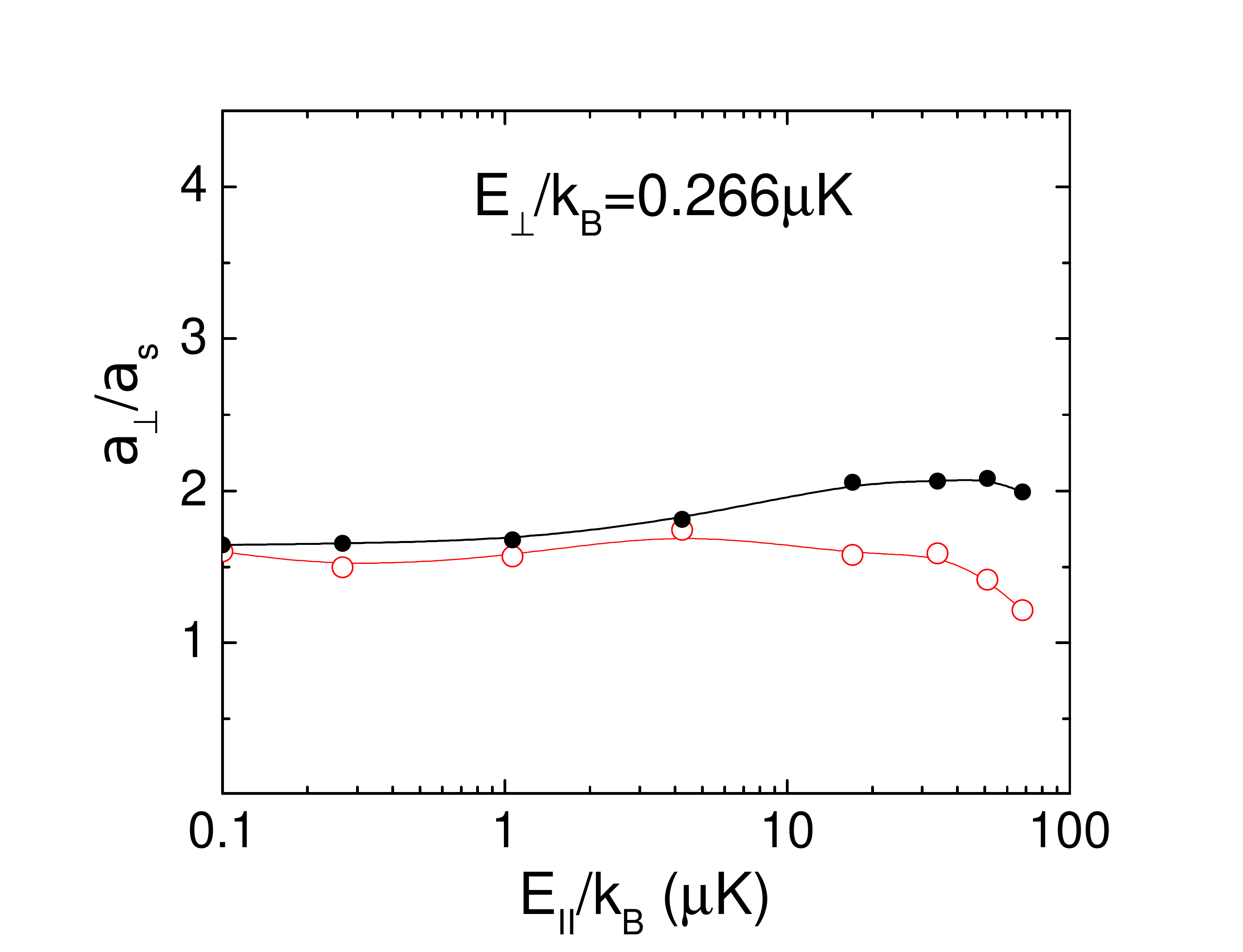}
\vspace{-.5cm}}
\parbox{0.95\textwidth}{
\includegraphics[width=0.45\textwidth,height=0.4\textwidth]{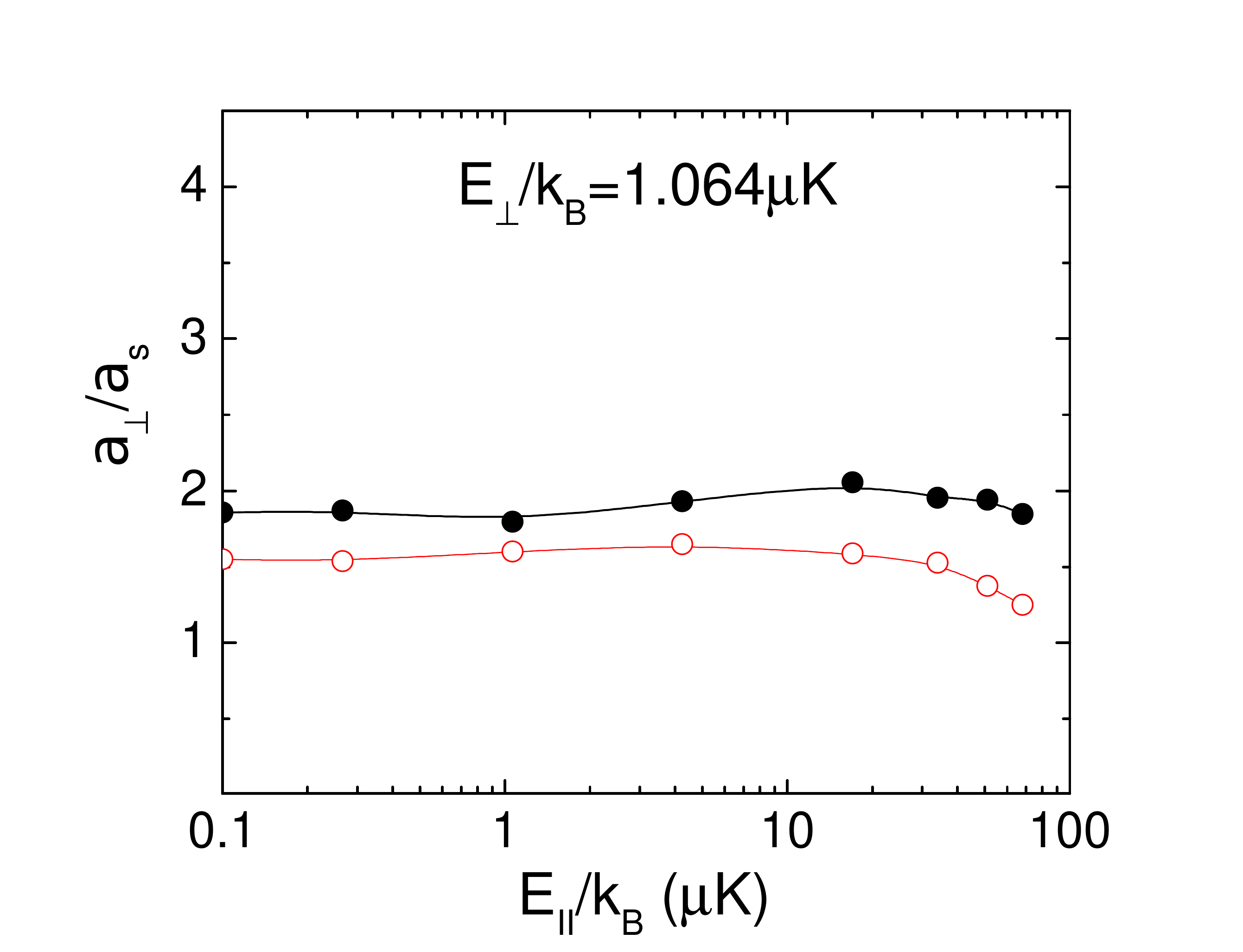}
\includegraphics[width=0.45\textwidth,height=0.4\textwidth]{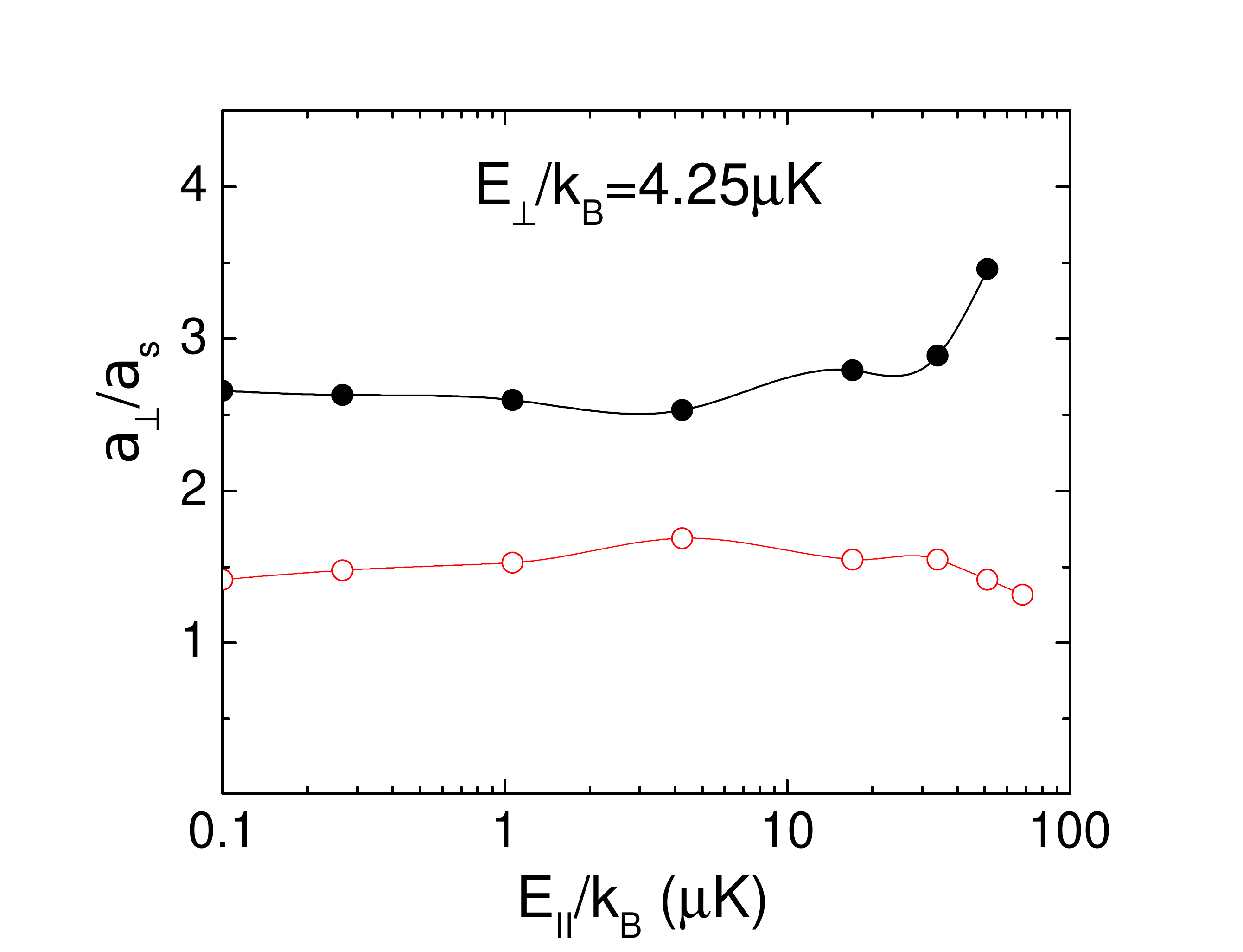}
\vspace{-.5cm}}
\parbox{0.95\textwidth}{
\includegraphics[width=0.45\textwidth,height=0.4\textwidth]{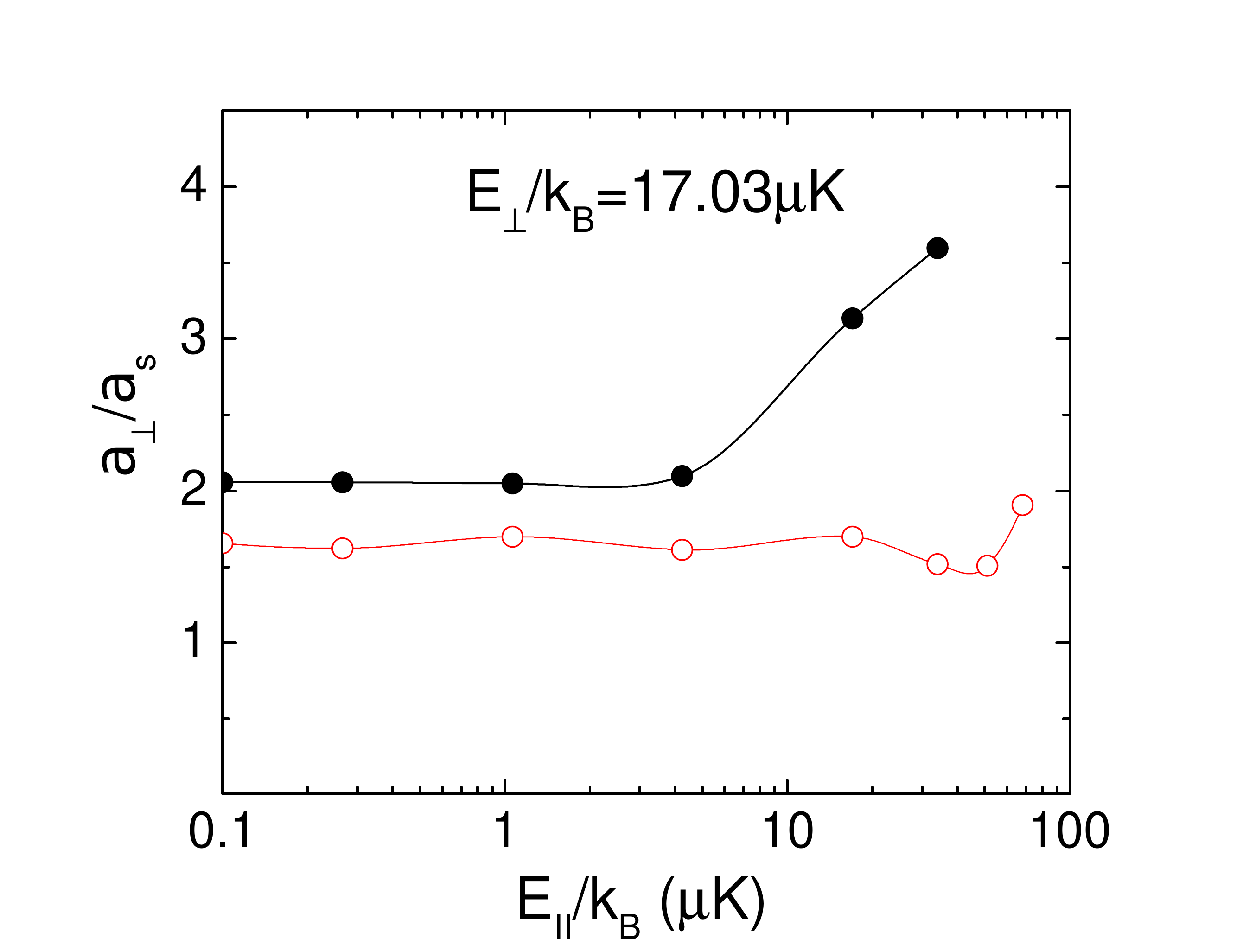}
\includegraphics[width=0.45\textwidth,height=0.4\textwidth]{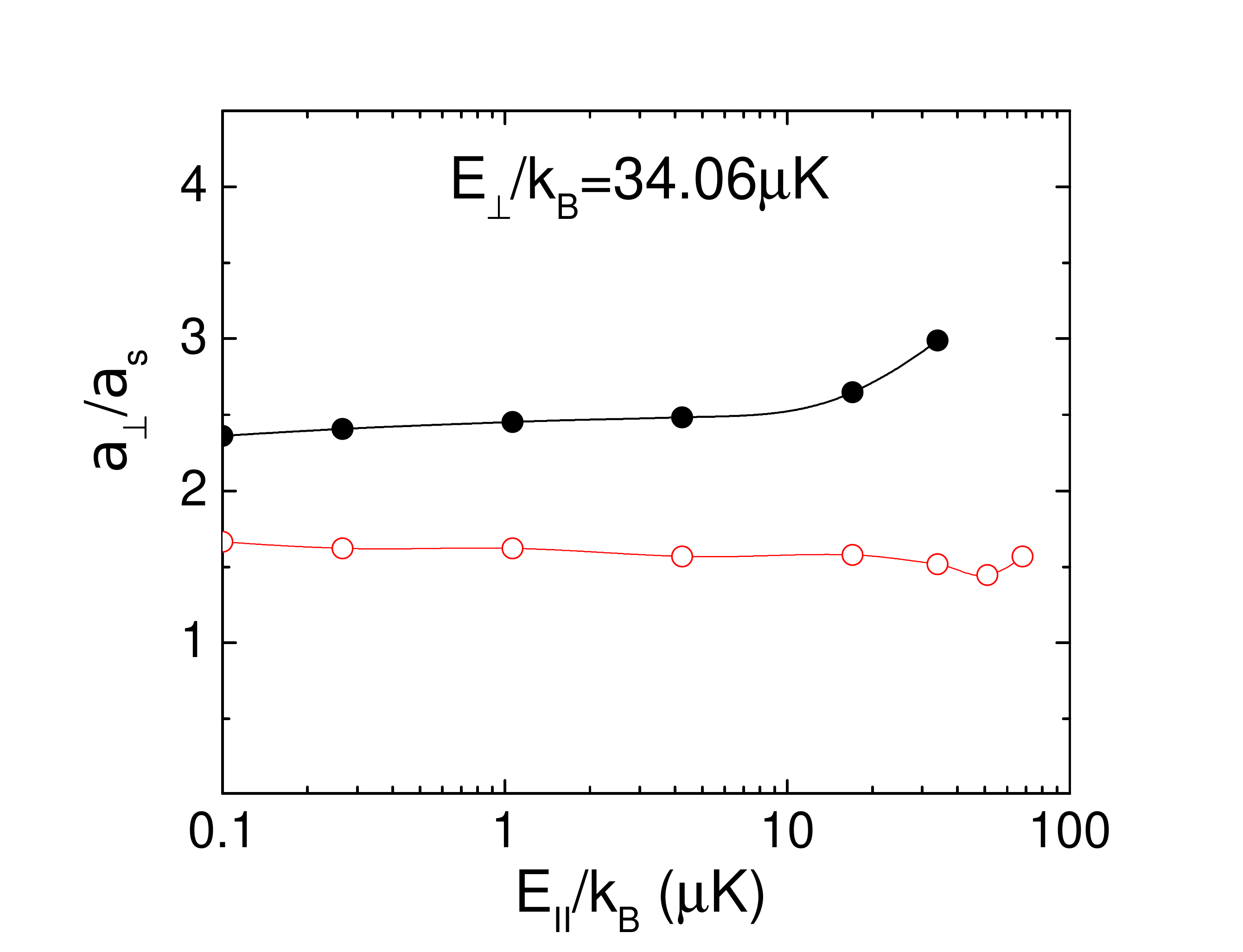}}
\vspace{0.5cm}
\caption{The dependence of the CIR position $a_{\perp}/a_s$ on the transversal $E_{\perp}$ and longitudinal $E_{\parallel}$ ion energies before the collision for the atomic trap with $\omega_{\perp}=0.03\,\omega^* =2\pi\times 11$ kHz. The (blue) arrow in the top left panel (i.e. $E_{\perp}=0$) indicates the atomic CIR position 1.4603 at the zero-energy limit~\cite{Olshanii1998}. Here the results of the calculations with the time-dependent Paul trap~(\ref{eq:Uion}) are given as black cycles connected by solid lines. The results obtained in the secular harmonic trap approximation~(\ref{eq:Usec}) are given as open cycles connected by (red) thin lines.
}
\label{fig:Fig6}
\end{figure*}

We proceed further with our analysis of the CIR position by investigating the impact of the time-dependent Paul trap with the confining potential defined in Eq.~(\ref{eq:Uion}). We have performed calculations for two confining atomic trap frequencies $\omega_{\perp}=0.02\omega^*$ and  $0.03\,\omega^*$ corresponding to $\omega_\perp = 2\pi\times 7.1$ kHz and $\omega_\perp = 2\pi\times 11$ kHz, respectively, for the pair $^6$Li/$^{174}$Yb$^+$. The results of the calculation of the CIR positions presented in Figs.~\ref{fig:Fig5} and~\ref{fig:Fig6} support the approximation of the secular trap in the rectangular domain $E_{\perp}\leq 1\mu$K, $E_{\|}\leq 5\mu$K.
Outside this domain, the positions $a_{\perp}/a_s$ calculated for the time-dependent Paul trap~(\ref{eq:Uion}) start deviate from the values calculated in the secular harmonic approach~(\ref{eq:Usec}). Herewith, the deviation increases with increasing ion energy. Moreover, out of the mentioned above energy domain we have found a different dependence of the CIR position on the transversal and longitudinal ion mean energies in the Paul trap. In the zero-energy limit for the ion initial energy ($E_{\perp}, E_{\parallel} \rightarrow 0$) we obtain the value for the CIR position coinciding with the result obtained earlier in the static ion approximation~\cite{MelezhikPRA16} as well as the same dependence on the ratio $R^*/a_{\perp}$ as in the case of the static ion approximation: with decreasing $a_{\perp}$ (increasing $\omega_{\perp}$) the position of the CIR shifts towards larger $a_{\perp}/a_s$ and deviates from the value 1.4603 (see the left top panels of Figs.~\ref{fig:Fig5} and ~\ref{fig:Fig6}).
However, for larger ion mean energies, outside the domain $E_{\perp}\leq 1\mu$K, $E_{\|}\leq 5\mu$K, the CIR position is not anymore captured by the static ion approximation.

The found effect of the ion motion in the radiofrequency field on the CIR position can be qualitatively interpreted as follows. Since the atom moves slowly relative the period of ion oscillations, it feels the averaged effective interaction with the ion. At a rather low ion energy (small amplitude of the ion vibrations), the ion motion only slightly corrects the aforementioned effective atom-ion interaction and slightly changes the relative collisional atom-ion energy, since we observe perfect reflection near the CIR position $a_{\perp}/a_s=1.46$ obtained in the approximation of a static ion. The CIR position (the point of perfect reflection) starts to deviate considerably from the value calculated in the  static ion approximation when the ion energy approaches the s-wave atom-ion collision threshold $E^*$.  Therefore, it is natural to assume here a considerable contribution of the p-wave to the scattering amplitude $f^+$. In the case of a comparable contribution of s- and p-waves, total transmission can be observed due to their interference - this is the so-called dual CIR predicted by J.I. Kim et. al.~\cite{KimMelSch}. However, the question of dual CIRs in hybrid atom-ion systems demands a separate analysis.

%----------------------------------------------------------------------------------------
%	SECTION 4
%----------------------------------------------------------------------------------------

\section{Conclusions}
\label{sec:conclusions}

We have investigated the conditions for the appearing of atom-ion CIRs in hybrid atom-ion systems. Our analysis has been done for the Li-atom confined in an optical trap situated within a linear Paul trap for a Yb-ion with the realistic parameters and takes into account the motion and micromotion of the ion. Such a choice for the atom-ion pair is motivated by the fact that it is very likely the only pair that permits to attain s-wave collisions in hybrid traps. The shifts of the CIRs due to the ion motion were calculated. We found that the energy of the ion provided by the oscillating radiofrequency fields can affect the resonance position substantially. However, in a broad range of the ion kinetic energies we found that the CIR position is stable even in presence of micromotion. These findings indicate that the intrinsic micromotion of the ion is not detrimental for the occurrence of the resonance and that its position can be controlled by the radiofrequency fields. This provides an additional mean for tuning atom-ion interactions in low spatial dimensions. It also indicates that experimental investigation of atom-ion CIRs in hybrid atom-ion systems is possible. Furthermore, we found a non-negligible probability (about 14\%) of forming an atom-ion molecule, thus indicating a pathway for producing such two-body compounds by controlling the confinement of the two atomic species.

The performed study of resonant collisions in the confined geometry of hybrid atom-ion traps represents an important advancement in the scattering physics of quantum systems in time-dependent traps. To this end,
we have adopted a quantum-semiclassical approach developed in Refs.~\cite{MelSchm,Melezhik2001,MelezhikCohen,MelSev}.
Our treatment can be applied for further studies of time-dependent problems, e.g. for simulations of two-qubit quantum gates~\cite{DoerkPRA10,StockPRL03} including the effects of micromotion \cite{NguyenPRA12}.
An extension of the method to hybrid atom-ion systems with comparable atom-ion mass ratios requires further work and it will be carried out in the future. In particular, the functional derivatives~(\ref{eq:funcDer}) will be taken into account.
Such studies will be important for understanding more deeply the reliance of the CIR position on the atom-ion mass ratio as well as the impact of the ion micromotion, namely whether a critical mass ratio does exist for the occurrence of CIRs. Indeed, in our study the small atom-ion mass ratio has played a major role to the fact that the micromotion is not too detrimental for the appearance of the CIR. Even more interesting will be to understand whether the s-wave regime can be attained in low spatial dimensions more favorably that in 3D for higher atom-ion mass ratios.

%----------------------------------------------------------------------------------------
%	SECTION 5
%----------------------------------------------------------------------------------------

\section{ACKNOWLEDGEMENTS}

The work was supported by the Russian Foundation for Basic Research, Grants No. 18-02-00673, by the Cluster of Excellence projects `The Hamburg Centre for Ultrafast Imaging' of the Deutsche Forschungsgemeinschaft (EXC 1074, Project No. 194651731) and `CUI: Advanced Imaging of Matter' of the Deutsche Forschungsgemeinschaft (EXC 2056, Project No. 390715994), and by the Polish National Science Center project 2014/14/M/ST2/00015.

%----------------------------------------------------------------------------------------
%	SECTION 6
%----------------------------------------------------------------------------------------

\appendix
\label{sec:appendix}

\section{Computational method}
\label{sec:numerics}

In order to integrate the semiclassical atom-ion equations of motion~(\ref{eq:SE}) and~(\ref{eq:Hamilton}), we applied the splitting-up method with a 2D discrete-variable representation (DVR)~\cite{Melezhik2012,Melezhik1997,Melezhik1998}.
For an accurate inclusion of the (strong) atom-ion potential~(\ref{eq:Vaireg}) in Eq.~(\ref{eq:SE}) at the moment of the resonant atom-ion collision, a tailored splitting-up procedure in the 2D-DVR representation has been developed, as we describe below. For the integration of the Hamilton equations of motion, which involve three considerably different scales of frequencies, namely $\Omega_{rf}$, $\omega_i$ as well as $\omega_{\perp}$ in the quantum mechanical average  $\langle\Psi(\vek{r}_a,t;\vek{r}_i)\vert V_{ai}(|\hat{\vek{r}}_a-\vek{r}_i(t)|)\vert\Psi(\vek{r}_a,t;\vek{r}_i)\rangle$, we employed the second order St\"ormer-Verlet method~\cite{Verlet}.

To begin with, the atom wavefunction $\Psi(\vek{r}_a,t)$ in the 2D-DVR is expanded as~\cite{Melezhik1997,Melezhik1998,MelBaye,Melezhik2016} (for the sake of simplicity, we omit hereafter the parametric dependence on the ion position)
\begin{align}
\label{eq:Psiexpansion}
\Psi\!(r_a,\Omega,t)=
\frac{1}{r_a}\sum_{j=1}^{N}f_j(\Omega)\,\psi_{j}(r_a,t)\,
\end{align}
with the 2D basis defined as
\begin{align}
\label{eq:fj}
f_j(\Omega)=\sum_{\nu=\{lm\}=1}^{N}Y_{\nu}(\Omega)(Y^{-1})_{\nu j}.
\end{align}
The latter is defined on an angular grid $\Omega_j =
(\theta_{j_{\theta}},\phi_{j_{\phi}})$ of $N$ grid points. The number $N$ is equal to the number of
basis functions in the expansion~(\ref{eq:Psiexpansion}) and the number of terms in
the definition~(\ref{eq:fj}). The coefficients $(Y^{-1})_{\nu j}$ in Eq.~(12) are the elements of the $N\times N$ matrix $Y^{-1}$
inverse to the matrix given by the values $Y_{j\nu}=Y_{\nu}(\Omega_j)$ of the polynomials $Y_{\nu}(\Omega)$ at the grid points $\Omega_j$. The construction of the 2D polynomials $Y_{\nu}(\Omega)$, which slightly deviate from the classical spherical harmonics for large $\nu\sim N$ that are orthogonal on the grid $\Omega_j$, is described in detail in Refs.~\cite{Melezhik1997,Melezhik1998,MelBaye,Melezhik2016}. We note that the coefficients $\psi_{j}(r_a,t)$ in Eq.~(\ref{eq:Psiexpansion}) define the values of the searching solution $\Psi(\vek{r}_a,t)$ at the points
of the angular grid $\Omega_j$: $\psi_{j}(r_a,t)=r_a \Psi(r_a,\Omega_{j},t)$.

With the 2D-DVR, the 3D Schr\"odinger equation~(\ref{eq:SE}) is thus approximated by the system of Schr\"odinger-like equations (hereafter $\hbar=1$)
\begin{equation}
\label{eq:psij}
i\frac{\partial}{\partial t}\bar{\psi}_j(r_a,t)=\sum_{j'}^N
[\hat H_{j}^{(0)}(r_a,t)\delta_{jj'}+\hat H_{jj'}^{(1)}(r_a)]\bar{\psi}_{j'}(r_a,t)\,\,,
\end{equation}
with the Hamiltonian consisting of the diagonal
\begin{equation}
\hat H_{j}^{(0)}(r_a) = -\frac{1}{2m_a}\frac{d ^{2}}{d r^{2}_a}
+\frac{1}{2}\omega_{\perp}^2r_a^2\sin^2\theta_j+V_{ai}(r_a,\Omega_j,\vek{r}_i(t))
\end{equation}
and offdiagonal
\begin{equation}
\hat H_{jj'}^{(1)}(r_a)=-\frac{1}{2m_a r_a^2}\frac{1}{\sqrt{\lambda_{j}\lambda_{j'}}}\sum_{\nu=\{lm\}=1}^N (Y^{-1}0_{j\nu}l(l+1)(Y^{-1})_{\nu j'}
\end{equation}
parts. Here $\lambda_j$ are the weights of the Gauss quadratures related with the (angular) grid $\Omega_j$~\cite{Melezhik2016} and $\bar{\psi}_j(r_a,t)=\sqrt{\lambda_j}\psi_j(r_a,t)$.

To integrate the system of equations~(\ref{eq:psij}) we use
the split-operator method that yields the propagation $\bar{\psi}_j(r_a,t_{n})\rightarrow \bar{\psi}_j(r_a,t_{n+1})$
with $t_n\rightarrow t_{n+1}=t_n+\Delta t$ according to
\begin{align}
\label{eq:psitn}
\bar{\psi}(t_{n}+\Delta t) &\approx
\exp\left(-\frac{i}{2}\Delta t \hat{H}^{(0)}\right) \exp(-i\Delta t\hat{H}^{(1)})\nonumber\\
\phantom{=}&\times\exp\left(-\frac{i}{2}\Delta t \hat{H}^{(0)}\right)\bar{\psi}(t_{n}) + O(\Delta t^3
).
\end{align}
The fact that the 2D-DVR [i.e, the functions~(\ref{eq:fj})] gives the diagonal
representation for $\hat{H}^{(0)}(r_a,t)$ and the $Y_{\nu}$-representation gives
the diagonal representations for $\hat{H}^{(1)}$, respectively, has been exploited for constructing an efficient computational algorithm.
Actually, for the first and the last steps of Eq.~(\ref{eq:psitn}), the exponential operators can be approximated according to
\begin{align}
&\exp\left[-\frac{i}{2}\Delta t \hat{H}_j^{(0)}(r_a,t_n)\right] \approx \left[1+\frac{i}{4}\Delta t
\hat{H}_j^{(0)}(r_a,t_n)\right]^{-1}\nonumber\\
\phantom{} &\times \left[1-\frac{i}{4}\Delta t \hat{H}_j^{(0)}(r_a,t_n)\right] + O(\Delta t^{2})\,,
\end{align}
Then, we have to solve $N$ independent second order differential equations
\begin{align}
& \left[1 + \frac{i}{4}\Delta t \hat{H}_j^{(0)}(r_a,t_n)\right]\bar{\psi}\left(t_{n} +
\frac{1}{2}\Delta t\right) = \nonumber\\
\phantom{} & \left[1 - \frac{i}{4}\Delta t\hat{H}_j^{(0)}(r_a,t_n)\right]\bar{\psi}(t_{n}).
\end{align}
The intermediate step in Eq.~(\ref{eq:psitn}) depending
on $\hat{H}_{jj'}^{(1)}(r_a)$ is treated in the basis $Y_{\nu}(\Omega_j)$, where the
matrix operator $\hat{H}_{\nu}^{(1)}$ is diagonal.
The transformation with help of the simple unitary matrix $S_{j
\nu}=\lambda_j^{1/2} Y_{j \nu}$ between 2D-DVR~(\ref{eq:fj}) and $Y_{\nu}$-representation provides the efficiency of the computational procedure~\cite{Melezhik1998,Melezhik2016}: computational time increases almost linearly with increasing number $N$ of basis functions.

Simultaneously to the forward in time propagation
$t_n\rightarrow t_{n+1}=t_n+\Delta t$ of the atom wave-packet $\psi_j(r_a,t_{n})\rightarrow \psi_j(r_a,t_{n+1})$ we integrate
the Hamilton equations~(\ref{eq:Hamilton}), which describe the dynamics of the ion in the Paul trap. To this end, we utilise the
second-order St\"ormer-Verlet method~\cite{Verlet}
\begin{equation}
\vek{p}_i^{(n+1/2)} = \vek{p}_i^{(n)} - \frac{\Delta t}{2}\frac{\partial}{\partial \vek{r}_i}H_i(\vek{p}_i^{(n+1/2)},\vek{r}_i^{(n)})\,,\nonumber
\end{equation}
\begin{align}
\vek{r}_i^{(n+1)} = \vek{r}_i^{(n)} + \frac{\Delta t}{2}\left\{\frac{\partial}{\partial \vek{r}_i}H_i(\vek{p}_i^{(n+1/2)},\vek{r}_i^{(n)})
\right.\nonumber \\
\left.+\frac{\partial}{\partial \vek{r}_i}H_i(\vek{p}_i^{(n+1/2)},\vek{r}_i^{(n+1)})\right\}\,,\nonumber
\end{align}
\begin{equation}
\vek{p}_i^{(n+1)} = \vek{p}_i^{(n+1/2)} - \frac{\Delta t}{2}\frac{\partial}{\partial \vek{r}_i}H_i(\vek{p}_i^{(n+1/2)},\vek{r}_i^{(n+1)}).
\end{equation}
with $\vek{p}_i^{(n)}=\vek{p}_i\left(t_n\right)$, $\vek{p}_i^{(n+1/2)}=\vek{p}_i\left(t_n+\frac{\Delta t}{2}\right)$, and $\vek{p}_i^{(n+1)}=\vek{p}_i\left(t_n+\Delta t\right)$ and the same definition for $\vek{r}_i^{(n)}$.

%----------------------------------------------------------------------------------------
%	SECTION 7
%----------------------------------------------------------------------------------------

\end{document}